\documentclass[a4paper,fleqn,usenatbib]{mnras}

\usepackage{txfonts}

\usepackage[T1]{fontenc}
\usepackage{ae,aecompl}
\usepackage{graphics}
\usepackage{longtable}

\usepackage{psfig}   
\usepackage{graphicx}
\usepackage{amssymb}
\usepackage{subfigure}

\title[Gas removal in SF regions]{The long-term effects of gas removal from hydrodynamic simulations of star formation}

\author[{\'C}alovi{\'c}, Klos, Hudson, {\emph et al}]{Aleksandra {\'C}alovi{\'c}$^{1,2}$\thanks{Authors made a joint contribution and each should be considered lead author.}, Katerina S. Klos$^{1,3\star}$, Robert B. Hudson$^{1\star}$, James E. Dale$^4$ and \newauthor
  \hspace*{-0.2cm} Richard J.~Parker\thanks{E-mail: R.Parker@sheffield.ac.uk}\thanks{Royal Society Dorothy Hodgkin Fellow}$^1$ \vspace*{0.1cm}\\
  $^1$Astrophysics Research Cluster, School of Mathematical and Physical Sciences, The University of Sheffield, Hicks Building, Sheffield, S3 7RH, UK \\
  $^2$School of Physics and Astronomy, University of Leicester, Leicester, LE1 7RH, UK\\
  $^3$School of Physics and Astronomy, University of St. Andrews, North Haugh, St Andrews, Fife KY16 9SS, UK\\
  $^4$Department of Peace and Conflict Research, Universitet Uppsala, Box 514, 751 20 Uppsala, Sweden \\}

\begin{document}

                             
\pagerange{\pageref{firstpage}--\pageref{lastpage}} \pubyear{2025}

\maketitle

\label{firstpage}

\begin{abstract}
The removal of gas left over from star formation has long been thought to dominate the dynamical evolution, and dissolution of star-forming regions. Feedback from massive stars from their stellar winds, photoionising radiation and supernovae is postulated to expel significant amounts of gas, altering the gravitational potential energy of the star-forming region and causing a supervirial expansion, which disperses the stars into the Galaxy on rapid timescales ($<$10\,Myr). The majority of previous work has utilised $N$-body simulations with a background potential to model the effects of gas removal. Here, we adopt a different approach where we take the end point of hydrodynamic simulations of star formation in which stars form with and without feedback from massive stars and then evolve the stars as $N$-body simulations. We also scale the velocities of the stars to various virial ratios, to mimic slower or faster removal of gas, and evolve these as additional $N$-body simulations. We find that the simulations where the stars inherit the velocities of the sink particles from the hydrodynamic simulations predominantly evolve more like a simulation in virial equilibrium, rather than the supervirial behaviour we would expect after gas removal. We see no significant differences in the dynamical evolution between the simulations where the stars inherit velocities directly from the hydrodynamical simulations and the simulations with (sub)virial velocities. This strongly suggests that gas removal by feedback processes does not lead to rapid expansion of star-forming regions, beyond the expansion caused by dynamical relaxation in star-forming regions.   
\end{abstract}


\section{Introduction}

Stars do not form in isolation, but rather in groups containing tens, to tens of thousands of stars \citep{Lada03,Bressert10}, and the stellar density in these groups exceeds that in the Galactic disc by several orders of magnitude \citep{Korchagin03}. Depending on the environmental conditions, these groups may coalesce into larger structures that are either gravitationally bound \citep[often referred to as `star clusters', e.g.][]{Zwart10,Kruijssen12b,Longmore14}, or unbound \citep[often referred to as `associations', e.g.][]{Blaauw64,Wright20,Wright22}.

In spite of this, star formation is an inherently inefficient process, with only a fraction of the gas in a Giant Molecular Cloud being converted into stars. This means that a significant constituent of the gravitational potential is the leftover gas. Many authors have posited that feedback (stellar winds, photoionising radiation, supernovae) will disperse the remaining gas, which changes the dynamical equilibrium of the stars, causing the star-forming region to expand and disperse \citep[e.g.][]{Tutukov78,Whitworth79,Lada84,Goodwin97a,Kroupa01a,Baumgardt07,Pfalzner15a,Shukirgaliyev18}.

The efficacy of gas removal/expulsion is the subject of intense debate in the literature. \citet{Bastian06} and \citet{Goodwin06} demonstrated that the velocity dispersions in star clusters are consistent with a supervirial state, i.e. they appear to be in the process of dynamical destruction from the gas removal. However, \citet{Gieles10} pointed out that these supervirial velocity dispersions are likely to be inflated due to hidden binaries, and \citet{Kruijssen12a} demonstrated that the local star formation efficiency in simulations is close to 100\,per cent, so any residual gas expulsion does not affect the virial state of the  stellar groups.

With the advent of \emph{Gaia} observations and complementary ground-based spectroscopic surveys, further studies have suggested that some star-forming regions appear to be undergoing expansion \citep{Bravi18,Kounkel18}, and these authors attribute this expansion to feedback from the massive stars.

A major problem with theoretical studies of the effects of gas removal on star-forming regions is that hydrodynamic simulations only follow the first few Myr of a GMC's evolution \citep[e.g.][]{Dale12b,Dale14,Bate14} and cannot follow the long-term effects of gas removal on the later dynamical evolution of the region, unless the sink particles/stars are extracted and run as an $N$-body or hybrid $N$-body/hydrodynamic simulation.

Some simulations \citep{Hubber13,Sills18,Karam23,Karam25} couple low-resolution SPH and $N$-body simulations to achieve a more self-consistent view of star formation, but these simulations do not yet model very massive star formation from the gas. 

An alternative approach is to take the output of hydrodynamical simulations in which feedback is present and then evolve the stars as an $N$-body simulation (\citealp{Moeckel10,Parker13a,Torniamenti22}; \citealp{Farias24,Bernard25}). The drawback of this approach is that the leftover gas is usually removed immediately prior to the $N$-body evolution (as in \citet{Moeckel10} and \citet{Parker13a}; however another approach \citep[e.g.][]{Farias24} is to replace the gas from the hydrodynamic simulation with a background potential in the subsequent $N$-body evolution). In principle, this should have a similar effect to the removal of a static background gas potential in $N$-body simulations \citep{Goodwin06,Adams06,Baumgardt07}, whereby the clusters immediately undergo a rapid expansion phase \citep{Pfalzner13c,Pfalzner15a}.

A possible difference between the approaches of including the gas as a background potential in an $N$-body simulation \citep{Goodwin06,Baumgardt07,Pfalzner13c} or modelling the later $N$-body evolution of stars formed in an SPH simulation \citep{Moeckel10,Parker13a,Parker17b} is that the stars in the latter type of simulation were not fully decoupled from the gas during their formation. This is especially important if the stars form under the influence of feedback, as the stars' velocities may symbiotically react to the removal of gas due to this feedback, rather than the stars' velocities reacting to the removal of the decoupled gas potential, which  in the $N$-body simulation is modelled as an additional massive particle \citep[though see][for an alternative approach]{Dinnbier20}. Therefore, removing the gas potential in the $N$-body simulation usually results in a drastic change in the dynamical evolution of the star particles \citep[e.g.][]{Lada84,Goodwin97a,Proszkow09}. 

In this paper, we  take the sink particles from five  SPH simulations \citep{Dale12b,Dale14} in which the stars have formed with feedback from massive stars, and we follow the subsequent $N$-body evolution of these particles. Whilst the parent simulations have been complemented by more recent simulations \citep[e.g.][]{Grudic21,Guszejnov22}, the more recent simulations explore a similar paramater space in terms of the GMC masses, radii, virial ratios and turbulence. The more recent simulations implement magnetic fields and a wider range of feedback mechanisms, but qualitatively form similar numbers of stars, with similar stellar densities, etc.

We follow the long-term $N$-body evolution of simulations in which we take the stars' velocities directly from the sink particle velocities in the SPH simulations of \citet{Dale12b,Dale14}, but discard the remaining gas which should mimic the effects of rapid gas removal. We then compare this to other runs of the same simulations, but where we scale the velocities to different virial ratios to determine how much the virial ratio affects the long-term evolution of these star-forming regions.

The paper is organised as follows. In Section~\ref{method} we descibe the SPH simulations, and the $N$-body simulations of the sink-particles that form in the SPH simulations. We present our results in Section~\ref{results}, and provide a discussion in Section~\ref{discuss}. We draw conclusions in Section~\ref{conclude}.

\section{Methods}
\label{method}

We take the sink particles from Smoothed Particle Hydrodynamic simulations of the formation of stars in Solar metallicity Giant Molecular Clouds first published in \citet{Dale12b} and \citet{Dale14}. \citet{Dale12b} and \citet{Dale14} present five simulations with different initial conditions; each different initial condtion has a control run in which the only physics is the conversion of gas into stars (sink particles), as well as a run where feedback from massive stars in the form of photoionisation radiation, and stellar winds, is implemented. In these simulations, the spatial resolution limit (i.e. sink particle radius) is 0.005\,pc.

Within the five pairs of simulations, the mass and radius of the initial cloud is varied, as is the initial virial ratio of the cloud. Simulations J and I were set up such that the gas cloud was bound (virial ratio $\alpha_{\rm cloud}^{\rm SPH} = 0.7$), whereas in simulations UF, UP and UQ the cloud was unbound (virial ratio $\alpha_{\rm cloud}^{\rm SPH} = 2.3$) -- the prefix `U' meaning `unbound'. Simulations J and I are both $M_{\rm cloud} = 10\,000$\,M$_\odot$, but have different radii (5\,pc and 10\,pc, respectively). Simulations, UF, UP and UQ each have different radii (10\,pc, 2.5\,pc and 5\,pc, respectively). Simulation UF has an initial cloud mass $M_{\rm cloud} = 30\,000$\,M$_\odot$ whereas UP and UQ have similar masses to simulations J and I ($M_{\rm cloud} = 10\,000$\,M$_\odot$).

  As reported in \citet{Dale12b,Dale14,Dale15b} and \citet{Parker13a} and \citet{Parker15a}, the main difference between the runs with and without feedback is that the control runs (no feedback) tend to form a more top-heavy Initial Mass Function with values for the exponent of the slope of the high-mass end of the IMF around $\alpha_3 \sim 1.5$, compared to a \citet{Salpeter55}-like slope ($\alpha_3 \sim 2.3$) for the run with feedback. The mass of the most massive star ranges between 54\,M$_\odot$ (simulation J) and 182\,M$_\odot$ (simulation UF) in the runs with no feedback, whereas for the runs with feedback the mass of the most massive star ranges between 29\,$M_\odot$ (simulation J) and 66\,M$_\odot$ (simulation I).  In addition,  the stellar density is higher in the control runs than in the runs with feedback.

  At the termination of the SPH runs (at around 2.5\,Myr after the formation of the first stars), we extract the sink particles and evolve them for a further 10\,Myr using the \texttt{kira} hermite $N$-body integrator within the \texttt{Starlab} environment \citep{Zwart99,Zwart01}.  Very few binary and multiple systems form during the simulations, but when they do they (and other close encounters) between stars are calculated directly using a reduced timestep within the block-timestepping used in \texttt{kira}.  We implement stellar and binary evolution  assuming Solar metallicity using the \texttt{SeBa} package \citep{Zwart96,Zwart12}, also within the \texttt{Starlab} environment.

  For snapshots of the spatial and density distributions of the stars and gas in these simulations at the end point of their evolution, we refer the interested reader to figs.~6~and~8 from \citet{Dale12b}, and figs.~8~and~9 from \citet{Dale14} (runs J and I), and fig.~1 from \citet{Dale13b} (runs UF, UP and UQ).

    In the runs without feedback, none of the gas particles are removed during the SPH simulation. Therefore, when we remove the gas before the $N$-body calculation, this change in gravitational potential represents the most extreme mass-loss event possible. The amount of gas removed from the simulations without feedback can be calculated by subtracting the total stellar mass ($M_{\rm region}$) from the cloud mass ($M_{\rm cloud}$); Run~J has the highest star-formation efficiency at 32\,per cent whereas Run~UF has the lowest star formation efficiency at 5\,per cent. In the classical picture of gas expulsion \citep[e.g.][]{Tutukov78,Whitworth79,Goodwin06,Parker17d}, we would expect all of these regions to become unbound following instantaneous gas removal.

  For the runs where the stars formed with feedback, the feedback can cause hot gas to be lost through holes in the cloud without destroying the entire cloud \citep{Lucas20}, and the combined effects of feedback also leads to gas-free, or gas-poor cavities around the stars. Therefore, for runs that formed with feedback, the removal of the remaining gas may not be expected to have as drastic an effect on the gravitational potential of the stellar groups, and therefore the later evolution of the region \citep[see also][]{Kruijssen12a}.

The aim of this paper is to investigate the effects of altering the initial virial ratio of the sink particle distributions on the long-term $N$-body evolution of the star-forming regions. Our default simulation runs are the pairs of simulations (control and dual feedback) in which we simply use the velocities of the sink particles as the initial conditions of the $N$-body simulations.

We then run an additional four pairs of simulations in which we force, or scale, the velocities of the sink particles to different initial virial ratios, defined as
\begin{equation}
  \alpha_{\rm vir} = \frac{T}{|\Omega|},
  \label{vir_eq}
\end{equation}
where $\alpha_{\rm vir} = 0.5$ is defined as virial equilibrium, and $\alpha_{\rm vir} < 0.5$ is subvirial, and $\alpha_{\rm vir} > 0.5$ is supervirial. Formally, a supervirial region is also said to be unbound if $\alpha_{\rm vir} > 1$.

In Eqn.~\ref{vir_eq} $T = \sum\limits_i T_i$ is the total kinetic energy of particles with individual kinetic energies $T_i$, which are given thus:
\begin{equation}
  T_i = \frac{1}{2}m_i|{\bf v}_i - {\bf v}_{\rm cl}|^2.
  \label{ke_eq}
\end{equation}
Here, ${\bf v}_i$ and ${\bf v}_{\rm cl}$ are the velocity vectors of the star and the centre of mass of the region, respectively. $|\Omega|$ is the sum of {\bf the} potential energies of the individual stars, $\sum\limits_i\Omega_i$, which are given by:
\begin{equation}
  \Omega_i = - \sum\limits_{i \not= j} \frac{Gm_im_j}{r_{ij}},
  \label{pe_eq}
\end{equation} 
where $m_i$ and $m_j$ are the masses of two stars and $r_{ij}$ is the distance between them.

The four different virial ratios we adopt are $\alpha_{\rm vir} = 0.01$, which results in very `cold' or slow velocities with respect to the gravitational potential, $\alpha_{\rm vir} = 0.3$ which gives slightly subvirial (`cool') velocities, $\alpha_{\rm vir} = 0.5$ (virial equilibrum) and $\alpha_{\rm vir} = 1.5$, which gives very supervirial (`hot', or fast) velocities.

We calculate several parameters to use to compare the different runs. The first is the half-mass radius, $r_H$, defined simply as the radius at the point where half the total stellar mass is enclosed. The total mass reduces throughout the simulation due to mass loss implemented by the \texttt{SeBa} stellar evolution code.

\begin{table*}
\caption[bf]{A summary of the five different pairs of smoothed particle hydrodynamics (SPH) simulations used as the input initial conditions of our $N$-body integrations. The values in the columns are: the simulation Run ID from \citet[][D12]{Dale12b}  or \citet[][D14]{Dale14}, whether feedback is implemented  in the SPH simulation, the paper reference,  the initial virial ratio of the original clouds ($\alpha_{\rm cloud}^{\rm SPH}$) (to distinguish bound from unbound clouds), the initial radius of the cloud in the SPH simulation ($R_{\rm cloud}$), the initial mass of the cloud ($M_{\rm cloud}$),   the number of stars that have formed at the end of the SPH simulation ($N_{\rm stars}$), the mass of this star-forming region ($M_{\rm region}$), the virial ratio of the stars that form ($\alpha_{\rm vir}^{\rm stars}$), the stellar half-mass radius of the region at the end of the SPH simulation ($r_H$), the median surface density of the stars at the end of the SPH simulation ($\Sigma_{\rm stars}$), the $\mathcal{Q}$-parameter of the stars at the end of the SPH simulation ($\mathcal{Q}_{\rm stars}$), and the mass segregation ratio of the stars at the end of the SPH simulation ($\Lambda_{\rm MSR}$).}
\begin{center}
  \begin{tabular}{|c|c|c|c|c|c|c|c|c|c|c|c|c|}
    Run ID & Feedback & Ref.  & $\alpha_{\rm cloud}^{\rm SPH}$ & $R_{\rm cloud}$ & $M_{\rm cloud}$ & $N_{\rm stars}$ & $M_{\rm region}$ & $\alpha_{\rm vir}^{\rm stars}$ & $r_H$ & $\Sigma_{\rm stars}$ &   $\mathcal{Q}_{\rm stars}$ & $\Lambda_{\rm MSR}$ \\

  \hline
  J & None & D12 & 0.7 & 5\,pc & 10\,000\,M$_\odot$ & 578 & 3207\,M$_\odot$ & 0.33 & 1.20\,pc & 5129\,stars\,pc$^{-2}$ &   0.49 & $0.78^{+1.03}_{-0.60}$    \vspace*{0.1cm}\\

  J & Dual & D14 & 0.7 & 5\,pc & 10\,000\,M$_\odot$ & 564 & 2186\,M$_\odot$ & 0.36 & 1.86\,pc &  56\,stars\,pc$^{-2}$ & 0.70 & $1.31^{+1.60}_{-1.01}$\\ 
  \hline
  I & None & D12 & 0.7 & 10\,pc & 10\,000\,M$_\odot$ & 186 & 1270\,M$_\odot$ & 0.30 & 0.70\,pc & 138\,stars\,pc$^{-2}$ & 0.72 &  $2.00^{+2.61}_{-1.40}$ \vspace*{0.1cm}\\
  I & Dual & D14 & 0.7 & 10\,pc & 10\,000\,M$_\odot$ & 132 & 766\,M$_\odot$ & 0.61 & 5.60\,pc & 6\,stars\,pc$^{-2}$ & 0.49 & $0.88^{+1.06}_{-0.75}$ \\
  \hline
  UF & None & D12 & 2.3 & 10\,pc & 30\,000\,M$_\odot$ & 66 & 1392\,M$_\odot$ & 0.22 & 1.49\,pc & 47\,stars\,pc$^{-2}$ & 0.77 &  $0.94^{+1.30}_{-0.48}$ \vspace*{0.1cm}\\
  UF & Dual & D14 & 2.3 & 10\,pc & 30\,000\,M$_\odot$ & 93 & 841\,M$_\odot$ & 2.42 & 9.74\,pc & 1\,stars\,pc$^{-2}$ & 0.49 & $1.02^{+1.21}_{-0.84}$ \\
  \hline
  UP & None & D12 & 2.3 & 2.5\,pc & 10\,000\,M$_\odot$ & 340 & 2718\,M$_\odot$ & 0.25 & 3.89\,pc & 264\,stars\,pc$^{-2}$ & 0.49 &  $0.92^{+1.19}_{-0.67}$ \vspace*{0.1cm}\\
  UP & Dual & D14 & 2.3 & 2.5\,pc & 10\,000\,M$_\odot$ & 343 & 1926\,M$_\odot$ & 0.24 & 3.98\,pc & 17\,stars\,pc$^{-2}$ & 0.64 & $0.94^{+1.19}_{-0.73}$ \\
  \hline
  UQ & None & D12 & 2.3 & 5\,pc & 10\,000\,M$_\odot$ & 48 & 723\,M$_\odot$ & 0.12 & 1.32\,pc & 10\,stars\,pc$^{-2}$ & 0.70 &  $0.80^{+1.10}_{-0.61}$ \vspace*{0.1cm}\\
  UQ & Dual & D14 & 2.3 & 5\,pc & 10\,000\,M$_\odot$ & 77 & 594\,M$_\odot$ & 0.34 & 7.51\,pc & 6\,stars\,pc$^{-2}$ & 0.45 & $0.85^{+0.99}_{-0.75}$ \\
  
\hline
\end{tabular}
\end{center}
\label{cluster_props}
\end{table*}

We also define the median stellar surface density as a measure of the evolution of stellar density that could be directly compared with observations of star-forming regions where the only information is in the two dimensional plane of the sky \citep[e.g.][]{Bressert10}. As a comparator between observations and simulations, it is generally robust against edge effects, extinction and membership uncertainty \citep{Parker12d}, and for substructured regions is more reliable than defining the central density, or density within the half-mass radius. To calculate it, we first determine the local surface density around each star as
\begin{equation}
  \Sigma = \frac{N - 1}{\pi D_N^2},
  \label{sig_eqn}
\end{equation}
where $D_N$ is the distance to the $N^{\rm th}$ nearest neighbour. The choice of $N$ is somewhat arbitrary, but ideally it should be greater than three or four in order to avoid density enhancements from binary or multiple systems, and less than ten to avoid ignoring subgroups which may have a higher stellar density than the rest of the star-forming region \citep[see e.g.][for a discussion]{Bressert10,Parker12d}. We adopt $N = 7$ in our calculations. We then determine $\Sigma$ for each star, and then plot the median for all stars at each timestep.

The structure of a star-forming region can be used to infer the likely initial density and/or virial ratio, with substructured regions at ages of several Myr or more implying supervirial velocities and/or low densities \citep{Parker12d,Parker14b,Wright14}. We quantify the substructure using the $\mathcal{Q}$-parameter \citep{Cartwright04,Cartwright09b,Lomax11}, defined as
\begin{equation}
  \mathcal{Q} = \frac{\bar{m}}{\bar{s}},
  \label{qpar_eqn}
\end{equation}
where $\bar{m}$ is the mean branch length of a minimum spanning tree that connects all the stars in the star-forming region by the shortest possible path where there are no closed loops, and $\bar{s}$ is the mean length of the complete graph, which links every star with every other star. When $\mathcal{Q} < 0.8$ the region is substructured, and when $\mathcal{Q}> 0.8$ the region is smooth and likely to be centrally concentrated. Efforts have been made to equate a particular type of geometry to aspecific combination of $\mathcal{Q}$, $\bar{m}$ and $\bar{s}$ \citep{Cartwright09b,Lomax18}, but usually non-idealised distributions (like those caused by dynamical evolution) make this highly non-trivial \citep{DaffernPowell20}.  

We quantify the degree of mass segregation in the simulations by using the $\Lambda_{\rm MSR}$ measure \citep{Allison09a}, which again uses minimum spanning trees to quantify the spatial distributions of stars relative to one another. Usually, mass segregation is defined as a difference between the distribution of the most massive stars compared to the distribution for all stars.

The mass segregation ratio,  $\Lambda_{\rm MSR}$, is defined as the average length of 100 sets of minimum spanning trees that connect $N_{\rm MST}$ stars, $\langle l_{\rm average} \rangle$,  divided by the length of the $N_{\rm MST}$ stars in a chosen subset, $l_{\rm subset}$, thus:   
\begin{equation}
  \Lambda_{\rm MSR} = {\frac{\langle l_{\rm average} \rangle}{l_{\rm subset}}} ^{+ {\sigma_{\rm 5/6}}/{l_{\rm subset}}}_{- {\sigma_{\rm 1/6}}/{l_{\rm subset}}}.
  \label{lambda_eqn}
\end{equation}
To quantify mass segregation using $\Lambda_{\rm MSR}$, we compare the minimum spanning tree of the $N_{\rm MST}$ most massive stars to the average of 100 minimum spanning trees of $N_{\rm MST}$ random stars. We begin by calculating  $\Lambda_{\rm MSR}$ for the $N_{\rm MST} = 4$ most massive stars,  and then repeat the calculation for the $N_{\rm MST} = 5$ most massive stars, and so on. This results in a large amount of data, and in the analysis we will present the results for the $N_{\rm MST} = 10$ most massive stars (though we have checked and the results are almost identical when we use the $N_{\rm MST} = 5$ most massive stars). A star-forming region is  mass segregated when $\Lambda_{\rm MSR}$ is significantly higher than unity, i.e.\,\,when the lower uncertainty is also higher than unity. We conservatively estimate the uncertainties as being the bound between the values for $\Lambda_{\rm MSR}$ for the average MST length for all stars 1/6 and 5/6 of the way through the ordered list of MST lengths. \\

We summarise the end-point of the SPH simulations in Table~\ref{cluster_props}, which shows the virial ratio of the stars, $\alpha_{\rm vir}^{\rm stars}$, at the point the gas is removed, the half-mass radius of the stars, $r_H$, the median surface density, $\Sigma_{\rm stars}$, the $\mathcal{Q}$-parameter and the $\Lambda_{\rm MSR}$ mass segregation ratio.

\begin{table*}
  \label{nbody_results}
\caption[bf]{Summary of the main results for all of the $N$-body runs of the SPH sink particle distributions. We indicate the run ID and whether feedback is implemented in the SPH calculations. We then show the virial ratio the stellar velocities are scaled to (`sinks' refers to simulations where the stars inherit the velocities from the SPH sink particles). We then show the parameters for each run at 0\,Myr (before $N$-body evolution) and at 10\,Myr (after $N$-body evolution). We show the half-mass radius ($r_H$), median stellar surface density, $\Sigma$, $\mathcal{Q}$-parameter and $\Lambda_{\rm MSR}$ mass segregation ratio.}
\begin{tabular}{|c|c|c|c|c|c|c|c|c|c|c|}
  \hline
    Run ID & Feedback & $\alpha_{\rm vir}$ & $r_{H, {\rm 0\,Myr}}$ &  $r_{H, {\rm 10\,Myr}}$ & $\Sigma_{\rm stars, 0\,Myr}$ &   $\Sigma_{\rm stars, 10\,Myr}$ & $\mathcal{Q}_{\rm stars, 0\,Myr}$ & $\mathcal{Q}_{\rm stars, 10\,Myr}$ & $\Lambda_{\rm MSR, 0\,Myr}$  & $\Lambda_{\rm MSR, 10\,Myr}$\\

    \hline
       &  & Sinks &  & 15.8\,pc &  & 0.38\,stars\,pc$^{-2}$ &  & 1.92 &   & $1.34^{+2.48}_{-0.80}$ \vspace*{0.1cm}  \\
       &  & 0.01 & & 14.4\,pc &  & 0.28\,stars\,pc$^{-2}$ &  & 1.67 &   & $0.001^{+0.018}_{-0.004}$ \vspace*{0.1cm}  \\
 J & None & 0.3 & 1.20\,pc & 13.9\,pc & 5129\,stars\,pc$^{-2}$ & 0.42\,stars\,pc$^{-2}$ & 0.49 & 1.95 & $0.78^{+1.03}_{-0.60}$  & $2.17^{+4.32}_{-1.16}$ \vspace*{0.1cm}  \\
       &  & 0.5 &  & 23.4\,pc &  & 0.24\,stars\,pc$^{-2}$ &  & 1.34 &  & $1.11^{+1.72}_{-0.76}$ \vspace*{0.1cm}  \\
       &  & 1.5 &  & 70.4\,pc &  & 0.03\,stars\,pc$^{-2}$ &  & 1.32 &   & $1.05^{+1.30}_{-0.86}$ \vspace*{0.1cm}  \\
    \hline
       &  & Sinks &  & 5.34\,pc &  & 2.5\,stars\,pc$^{-2}$ &  & 1.50 &   & $0.81^{+1.29}_{-0.49}$ \vspace*{0.1cm}  \\
       &  & 0.01 &  & 3.05\,pc &  & 3.4\,stars\,pc$^{-2}$ & & 1.95 &   & $1.02^{+2.01}_{-0.55}$ \vspace*{0.1cm}  \\  
 J & Dual & 0.3 & 1.86\,pc & 3.05\,pc & 56\,stars\,pc$^{-2}$ & 6.2\,stars\,pc$^{-2}$ & 0.70 & 1.66 & $1.31^{+1.60}_{-1.01}$  & $2.82^{+4.76}_{-1.59}$ \vspace*{0.1cm}  \\
       &  & 0.5 &  & 5.85\,pc &  & 1.7\,stars\,pc$^{-2}$ &  & 1.26 &   & $0.71^{+1.06}_{-0.47}$ \vspace*{0.1cm}  \\
       &  & 1.5 &  & 24.4\,pc & & 0.14\,stars\,pc$^{-2}$ &  & 1.06 &   & $0.93^{+1.17}_{-0.73}$ \vspace*{0.1cm}  \\
    \hline
    \hline
      &  & Sinks &  & 8.72\,pc &  & 0.27\,stars\,pc$^{-2}$ &  & 1.39 &  & $3.40^{+5.01}_{-2.20}$ \vspace*{0.1cm}  \\
      &  & 0.01 &  & 15.8\,pc &  & 0.09\,stars\,pc$^{-2}$ &  & 1.46 &   & $0.40^{+0.73}_{-0.25}$ \vspace*{0.1cm}  \\  
I & None & 0.3 & 0.70\,pc & 14.0\,pc & 138\,stars\,pc$^{-2}$ & 0.15\,stars\,pc$^{-2}$ & 0.72 & 1.21 & $2.00^{+2.61}_{-1.40}$  & $1.77^{+2.36}_{-1.31}$ \vspace*{0.1cm}  \\
      &  & 0.5 &  & 11.4\,pc &  & 0.14\,stars\,pc$^{-2}$ &  & 1.14 &   & $2.26^{+3.18}_{-1.59}$ \vspace*{0.1cm}  \\
      &  & 1.5 &  & 40.7\,pc &  & 0.02\,stars\,pc$^{-2}$ & & 0.96 &   & $1.08^{+1.35}_{-0.89}$ \vspace*{0.1cm}  \\
    \hline
      &  & Sinks &  & 26.2\,pc &  & 0.03\,stars\,pc$^{-2}$ & & 0.60 &  & $0.95^{+1.18}_{-0.78}$ \vspace*{0.1cm}  \\
      &  & 0.01 &  & 7.75\,pc & & 0.37\,stars\,pc$^{-2}$ & & 1.15 &  & $0.85^{+1.33}_{-0.56}$ \vspace*{0.1cm}  \\  
I & Dual & 0.3 & 5.60\,pc & 18.0\,pc & 6\,stars\,pc$^{-2}$ & 0.20\,stars\,pc$^{-2}$ & 0.49 & 0.75 & $0.88^{+1.06}_{-0.75}$  & $1.23^{+1.49}_{-0.98}$ \vspace*{0.1cm}  \\
      &  & 0.5 &  & 22.0\,pc & & 0.17\,stars\,pc$^{-2}$ &  & 0.73 &  & $1.08^{+1.42}_{-0.84}$ \vspace*{0.1cm}  \\
      &  & 1.5 &  & 35.1\,pc &  & 0.03\,stars\,pc$^{-2}$ & & 0.72 &   & $1.27^{+1.54}_{-1.03}$ \vspace*{0.1cm}  \\
    \hline
    \hline
       &  & Sinks &  & 20.5\,pc &  & 0.02\,stars\,pc$^{-2}$ &  & 1.02 &   & $1.60^{+2.04}_{-1.17}$ \vspace*{0.1cm}  \\
       &  & 0.01 &  & 17.4\,pc &  & 0.03\,stars\,pc$^{-2}$ &  & 1.10 &   & $0.93^{+1.23}_{-0.65}$ \vspace*{0.1cm}  \\  
UF & None & 0.3 & 1.49\,pc & 20.2\,pc & 47\,stars\,pc$^{-2}$ & 0.02\,stars\,pc$^{-2}$ & 0.77 & 1.04 & $0.94^{+1.30}_{-0.48}$  & $1.71^{+2.80}_{-1.26}$ \vspace*{0.1cm}  \\
       &  & 0.5 &  & 16.5\,pc &  & 0.04\,stars\,pc$^{-2}$ &  & 0.99 &   & $1.43^{+1.82}_{-1.09}$ \vspace*{0.1cm}  \\
       &  & 1.5 &  & 47.5\,pc &  & 0.006\,stars\,pc$^{-2}$ &  & 0.93 &   & $1.06^{+1.30}_{-0.84}$ \vspace*{0.1cm}  \\
    \hline
       &  & Sinks &  & 37.7\,pc & & 0.03\,stars\,pc$^{-2}$ &  & 0.73 &   & $0.98^{+1.17}_{-0.81}$ \vspace*{0.1cm}  \\
       &  & 0.01 &  & 12.7\,pc & & 0.24\,stars\,pc$^{-2}$ &  & 1.15 &   & $0.93^{+1.64}_{-0.64}$ \vspace*{0.1cm}  \\
UF & Dual & 0.3 & 9.74\,pc & 18.6\,pc & 1\,stars\,pc$^{-2}$ & 0.09\,stars\,pc$^{-2}$ & 0.49 & 0.79 & $1.02^{+1.21}_{-0.84}$  & $1.45^{+1.91}_{-1.10}$ \vspace*{0.1cm}  \\
       &  & 0.5 &  & 21.2\,pc &  & 0.10\,stars\,pc$^{-2}$ &  & 0.70 &   & $1.43^{+1.70}_{-1.19}$ \vspace*{0.1cm}  \\
       &  & 1.5 &  & 28.0\,pc &  & 0.03\,stars\,pc$^{-2}$ &  & 0.83 &  & $1.38^{+1.87}_{-1.12}$ \vspace*{0.1cm}  \\
    \hline
    \hline
       &  & Sinks &  & 19.6\,pc &  & 0.18\,stars\,pc$^{-2}$ &  & 1.40 &   & $0.19^{+0.47}_{-0.12}$ \vspace*{0.1cm}  \\
       &  & 0.01 & & 65.1\,pc & & 0.29\,stars\,pc$^{-2}$ &  & 1.60 &   & $1.01^{+2.50}_{-0.49}$ \vspace*{0.1cm}  \\  
UP & None & 0.3 & 3.89\,pc & 19.8\,pc & 264\,stars\,pc$^{-2}$ & 0.17\,stars\,pc$^{-2}$ & 0.49 & 1.47 & $0.92^{+1.19}_{-0.67}$  & $0.50^{+0.90}_{-0.33}$ \vspace*{0.1cm}  \\
       &  & 0.5 &  & 24.0\,pc &  & 0.07\,stars\,pc$^{-2}$ &  & 1.20 &   & $1.31^{+2.02}_{-0.95}$ \vspace*{0.1cm}  \\
       &  & 1.5 &  & 60.5\,pc &  & 0.02\,stars\,pc$^{-2}$ &  & 1.09 &   & $0.87^{+1.14}_{-0.69}$ \vspace*{0.1cm}  \\
    \hline
       &  & Sinks &  & 15.7\,pc &  & 0.23\,stars\,pc$^{-2}$ &  & 1.01 &   & $0.78^{+1.08}_{-0.57}$ \vspace*{0.1cm}  \\
       &  & 0.01 &  & 5.36\,pc &  & 2.5\,stars\,pc$^{-2}$ & & 2.31 &   & $0.19^{+0.71}_{-0.10}$ \vspace*{0.1cm}  \\  
UP & Dual & 0.3 & 3.98\,pc & 13.7\,pc & 17\,stars\,pc$^{-2}$ & 0.29\,stars\,pc$^{-2}$ & 0.64 & 1.45 & $0.94^{+1.19}_{-0.73}$  & $1.10^{+1.68}_{-0.79}$ \vspace*{0.1cm}  \\
       &  & 0.5 &  & 22.2\,pc & & 0.12\,stars\,pc$^{-2}$ &  & 1.29 &   & $0.36^{+0.49}_{-0.27}$ \vspace*{0.1cm}  \\
       &  & 1.5 &  & 48.1\,pc &  & 0.03\,stars\,pc$^{-2}$ &  & 1.09 &   & $0.74^{+0.94}_{-0.58}$ \vspace*{0.1cm}  \\
   \hline
\end{tabular}
  \end{table*}

  \begin{table*}
    \begin{tabular}{|l|l|l|l|l|l|l|l|l|l|l|}
      \textbf{Table~\ref{nbody_results}} - continued\\
      \hline
    Run ID & Feedback & $\alpha_{\rm vir}$ & $r_{H, {\rm 0\,Myr}}$ &  $r_{H, {\rm 10\,Myr}}$ & $\Sigma_{\rm stars, 0\,Myr}$ &   $\Sigma_{\rm stars, 10\,Myr}$ & $\mathcal{Q}_{\rm stars, 0\,Myr}$ & $\mathcal{Q}_{\rm stars, 10\,Myr}$ & $\Lambda_{\rm MSR, 0\,Myr}$  & $\Lambda_{\rm MSR, 10\,Myr}$\\
    \hline
     &  & Sinks &  & 13.7\,pc &  & 0.02\,stars\,pc$^{-2}$ &  & 0.93 &   & $0.93^{+1.14}_{-0.70}$ \vspace*{0.1cm}  \\
     &  & 0.01 &  & 15.2\,pc &  & 0.05\,stars\,pc$^{-2}$ &  & 0.68 &   & $1.00^{+1.00}_{-0.01}$ \vspace*{0.1cm}  \\  
   UQ & None & 0.3 & 1.32\,pc & 19.1\,pc & 10\,stars\,pc$^{-2}$ & 0.03\,stars\,pc$^{-2}$ & 0.70 & 0.94 & $0.80^{+1.10}_{-0.61}$  & $1.20^{+2.02}_{-0.87}$ \vspace*{0.1cm}  \\
     &  & 0.5 &  & 26.5\,pc &  & 0.009\,stars\,pc$^{-2}$ &  & 0.87 &   & $1.30^{+1.56}_{-1.06}$ \vspace*{0.1cm}  \\
     &  & 1.5 &  & 33.7\,pc &  & 0.003\,stars\,pc$^{-2}$ &  & 0.96 &   & $1.41^{+1.75}_{-1.13}$ \vspace*{0.1cm}  \\
    \hline
         &  & Sinks & & 32.8\,pc &  & 0.03\,stars\,pc$^{-2}$ & & 0.68 &  & $1.52^{+1.79}_{-1.28}$ \vspace*{0.1cm}  \\
     &  & 0.01 & & 11.1\,pc &  & 0.1\,stars\,pc$^{-2}$ &  & 1.15 &  & $2.13^{+3.40}_{-1.51}$ \vspace*{0.1cm}  \\  
    UQ & Dual & 0.3 & 7.51\,pc & 28.1\,pc & 6\,stars\,pc$^{-2}$ & 0.04\,stars\,pc$^{-2}$ & 0.45 & 0.72 & $0.85^{+0.99}_{-0.72}$  & $1.16^{+1.53}_{-0.95}$ \vspace*{0.1cm}  \\
    &  & 0.5 &  & 33.5\,pc &  & 0.03\,stars\,pc$^{-2}$ &  & 0.66 &   & $1.28^{+1.50}_{-1.08}$ \vspace*{0.1cm}  \\
     &  & 1.5 &  & 51.1\,pc &  & 0.009\,stars\,pc$^{-2}$ &  & 0.69 &   & $1.00^{+1.16}_{-0.85}$ \vspace*{0.1cm}  \\
\hline
\end{tabular}
\end{table*}

\section{Results}
\label{results}

In this section we present various measures of the evolution of the sink particles as a pure $N$-body simulation, as we use as an input the end-point of the SPH simulations from \citet{Dale12b} and \citet{Dale14}. In each subsection we will first describe the evolution without any feedback from the massive stars, and then describe the evolution of the runs where photoionising radiation and stellar winds acted upon the clouds in the SPH simulation.  For brevity, and to facilitate better readability of the figures, we do not show all five pairs of simulations, but instead focus on three (J, I and UF); the results for the remaining two (UP and UQ) are qualitatively similar, and we include the main results from these runs in Table~\ref{nbody_results}.

The $N$-body evolution of a star-forming region is often framed in terms of the expected relaxation time, $t_{\rm rel}$ of the system,
  \begin{equation}
    t_{\rm rel} = t_{\rm cross}\frac{N}{8{\rm ln}N},
    \label{eqn:relax}
  \end{equation}
  where $N$ is the total number of stars and $t_{\rm cross}$ is the crossing time of the system, usually approximated by
  \begin{equation}
    t_{\rm cross} = \frac{R}{\sigma},
  \end{equation}
  where $R$ is the radius of the star-forming region and $\sigma$ is the velocity dispersion. For two star-forming regions with similar numbers of stars and similar radii, a region with a higher velocity dispersion should relax on a shorter timescale. However, whilst Eqn.~\ref{eqn:relax} may provide insights into the dynamical relaxation of populous, centrally concentrated star clusters \citep{Binney87}, it routinely fails to predict the timescales for e.g. the \citet{Spitzer69} instability, or for mass segregation to develop \citep{Bonnell98,Allison10,Parker16c}.

  When we refer to relaxation times in our analysis, we note that the dynamical timescale of importance is set by the local stellar density (which we approximate using the two-dimensional equivalent median surface density, as defined in Eqn.~\ref{sig_eqn}), not the average, or central density, both of which underpredict the degree of dynamical evolution that takes place. A region can evolve (relax) by all of the stars interacting with each other (in which case the structure would be erased and the $\mathcal{Q}$-parameter will be high), or the substructures within a region can move apart and the region would still appear substructured ($\mathcal{Q}<0.8$), even if significant dynamical relaxation had occurred within the substructures \citep{Parker18b}.

\subsection{Evolution of the stellar mass}

\begin{figure*}
  \begin{center}
    \setlength{\subfigcapskip}{10pt}
    \hspace*{-1.3cm} \subfigure[Run J]{\label{N_mass-a}\rotatebox{270}{\includegraphics[scale=0.28]{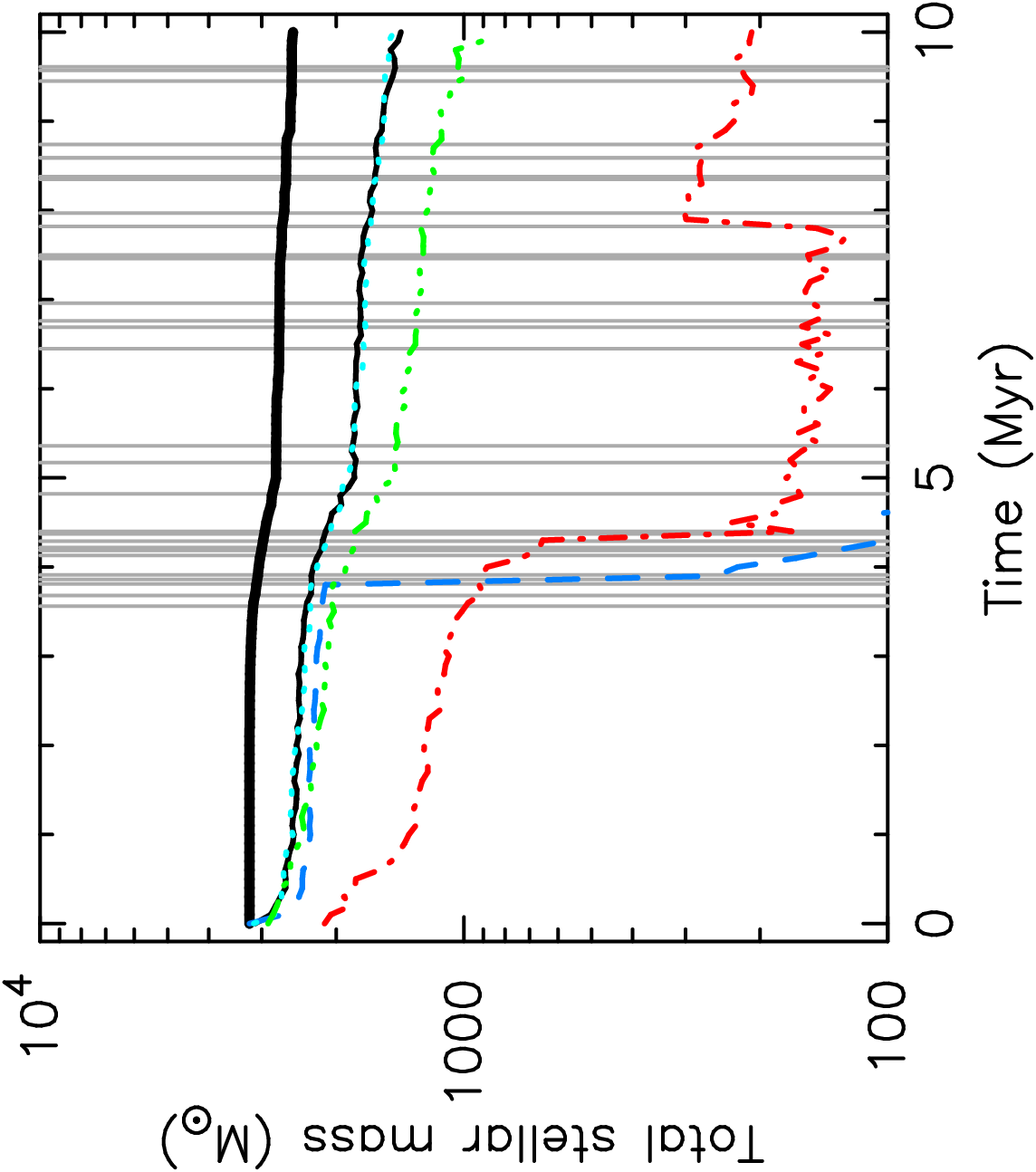}}}
    \hspace*{-0cm} \subfigure[Run I]{\label{N_mass-b}\rotatebox{270}{\includegraphics[scale=0.28]{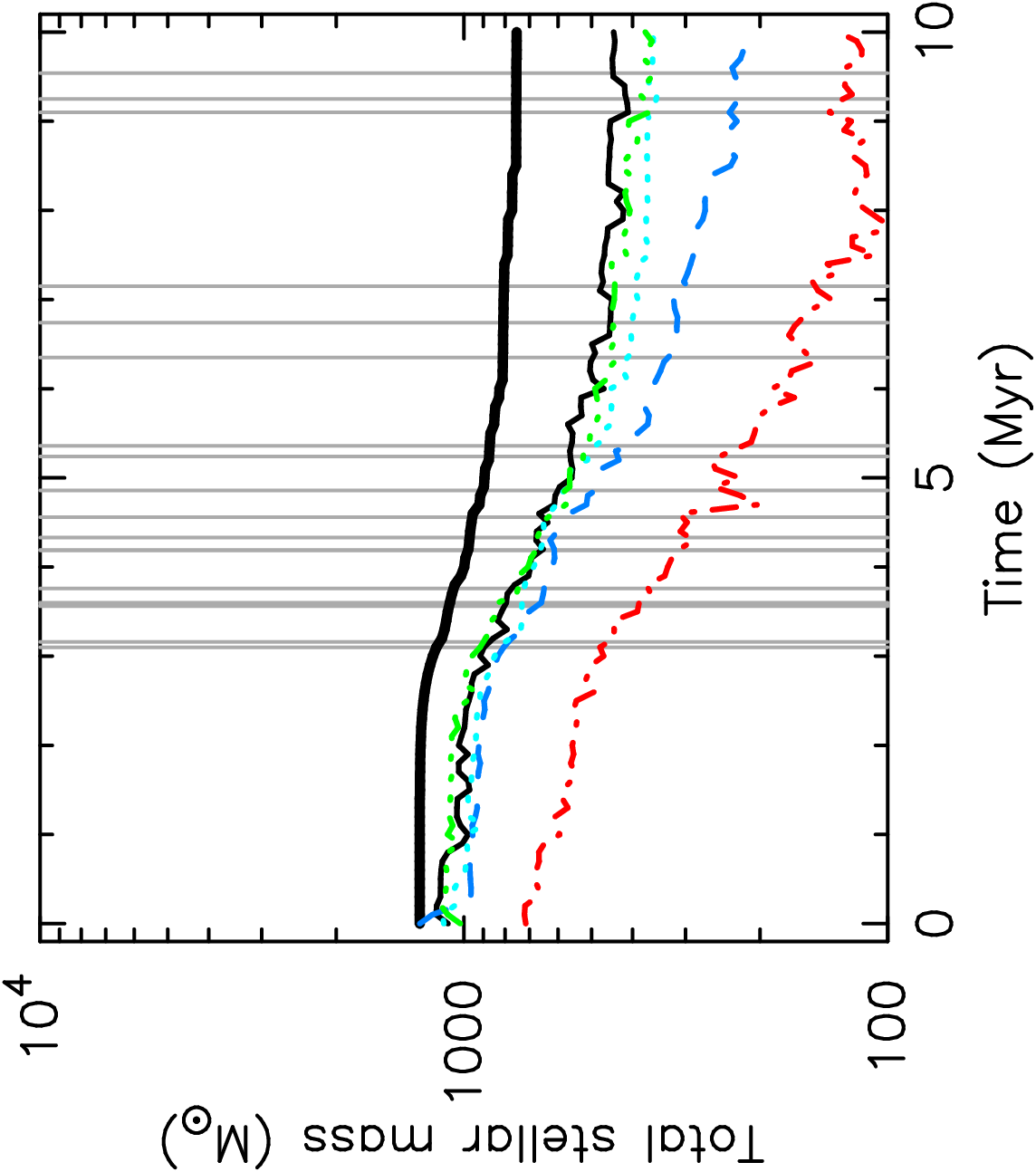}}}
    \hspace*{-0cm} \subfigure[Run UF]{\label{N_mass-c}\rotatebox{270}{\includegraphics[scale=0.28]{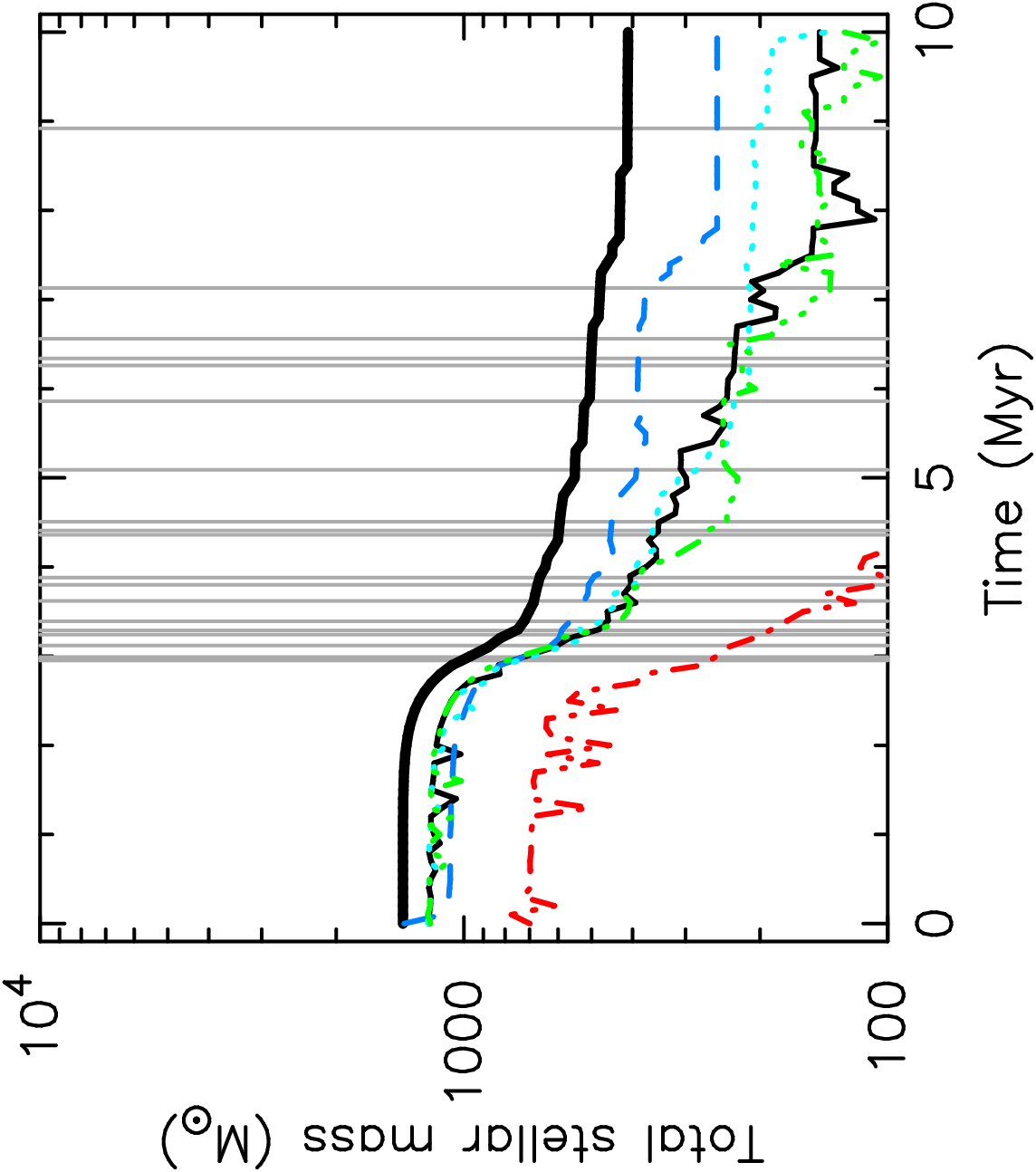}}}
    
\caption[bf]{The evolution of total stellar mass in three of our control run simulations in which feedback is not implemented. The thick black lines show the total mass, which decreases due to mass loss from the stars from their stellar evolution. The vertical lines indicate the time at which the first stage of post-main sequence evolution for each star that does leave the main sequence occurs.  The thin lines show the evolution of the bound star mass  for the versions of the simulations with different velocity scalings. The thin solid black lines are the original (sink particle) velocities, dashed dark blue lines are for $\alpha_{\rm vir} = 0.01$, dotted cyan lines are for $\alpha_{\rm vir} = 0.3$, dot-dot-dot-dashed green lines are for $\alpha_{\rm vir} = 0.5$ and dot-dashed red lines are for $\alpha_{\rm vir} = 1.5$.} 
\label{N_mass}
  \end{center}
\end{figure*}

In Fig.~\ref{N_mass} we show the evolution of the total stellar mass for the $N$-body simulations of the runs that formed stars without feedback, shown by the thicker (top) line in all panels. As detailed in \citet{Dale12b} and \citet{Parker13a}, the absence of feedback in these simulations allows very massive stars to form, making a slightly top-heavy IMF. This means that many of the stars undergo mass-loss (implemented by the \texttt{SeBa} package) and begin to leave the main sequence from 3\,Myr onwards; in Fig.~\ref{N_mass} we plot a vertical line at the time each star begins its first post-main sequence  phase. This is most pronounced in simulation UF (Fig.~\ref{N_mass-c}), where significant mass-loss occurs after 3\,Myr. 

\begin{figure*}
  \begin{center}
    \setlength{\subfigcapskip}{10pt}
    \hspace*{-1.3cm} \subfigure[Run J]{\label{B_mass-a}\rotatebox{270}{\includegraphics[scale=0.275]{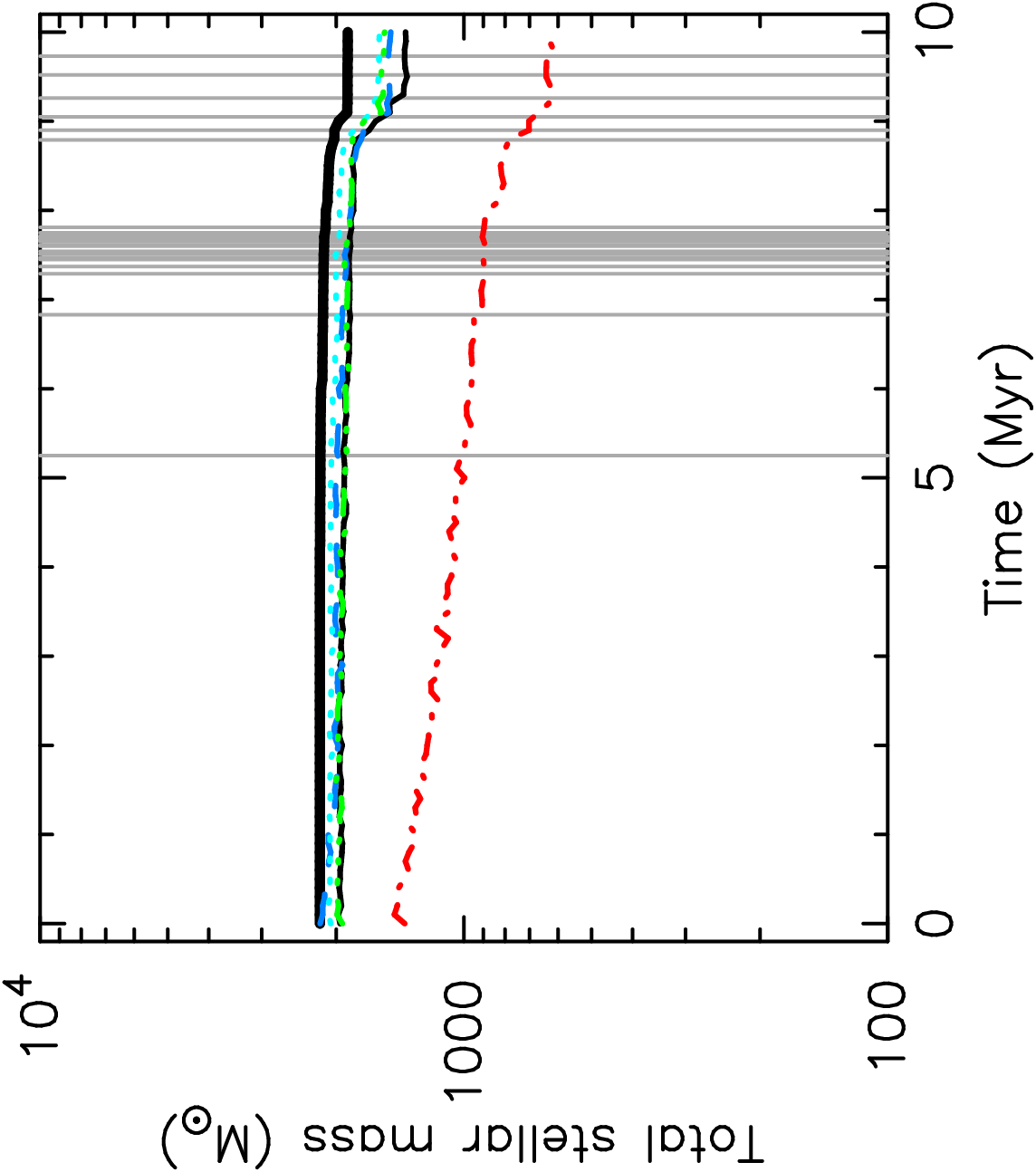}}}
    \hspace*{0.1cm} \subfigure[Run I]{\label{B_mass-b}\rotatebox{270}{\includegraphics[scale=0.275]{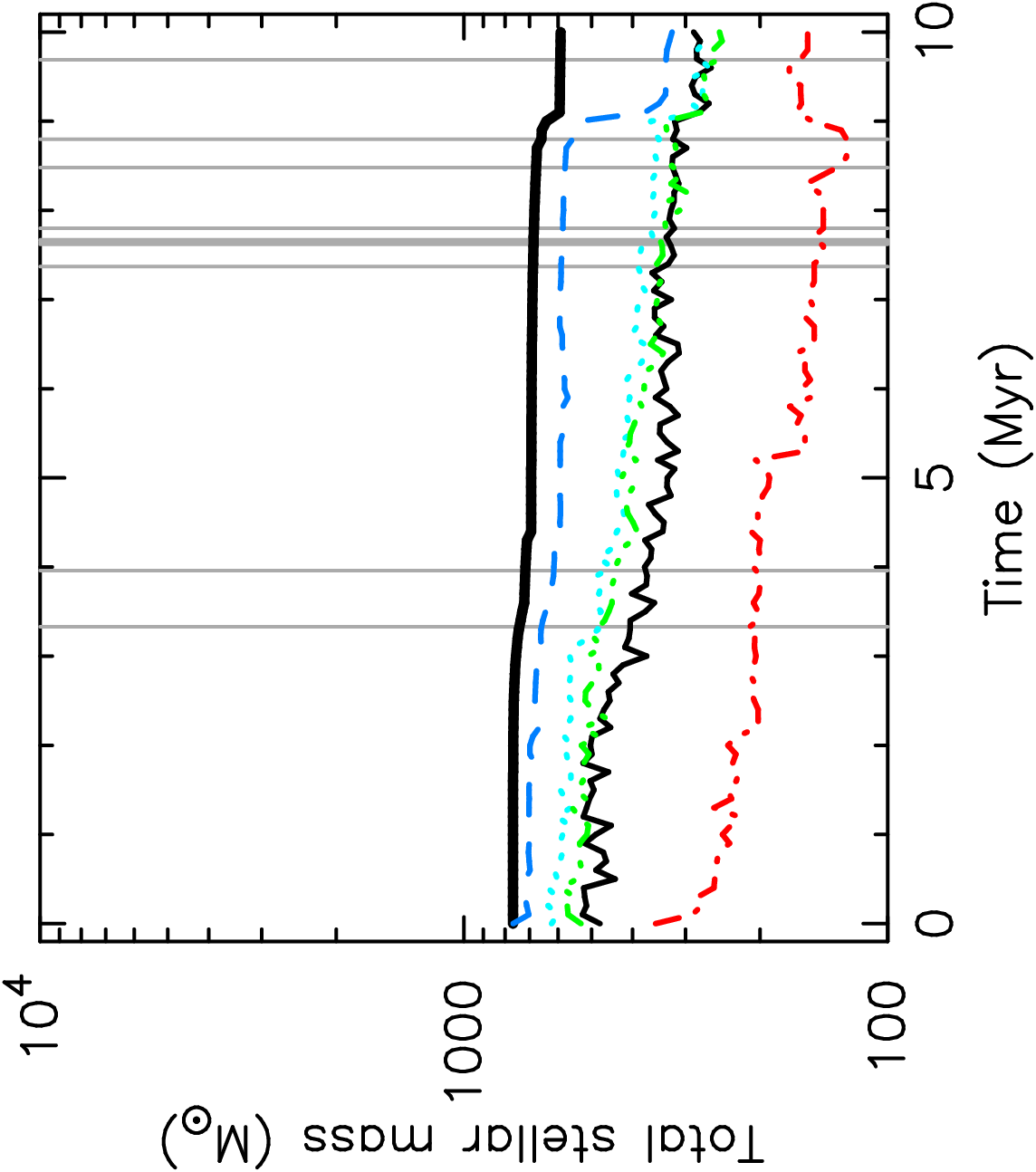}}}
    \hspace*{0.1cm} \subfigure[Run UF]{\label{B_mass-c}\rotatebox{270}{\includegraphics[scale=0.275]{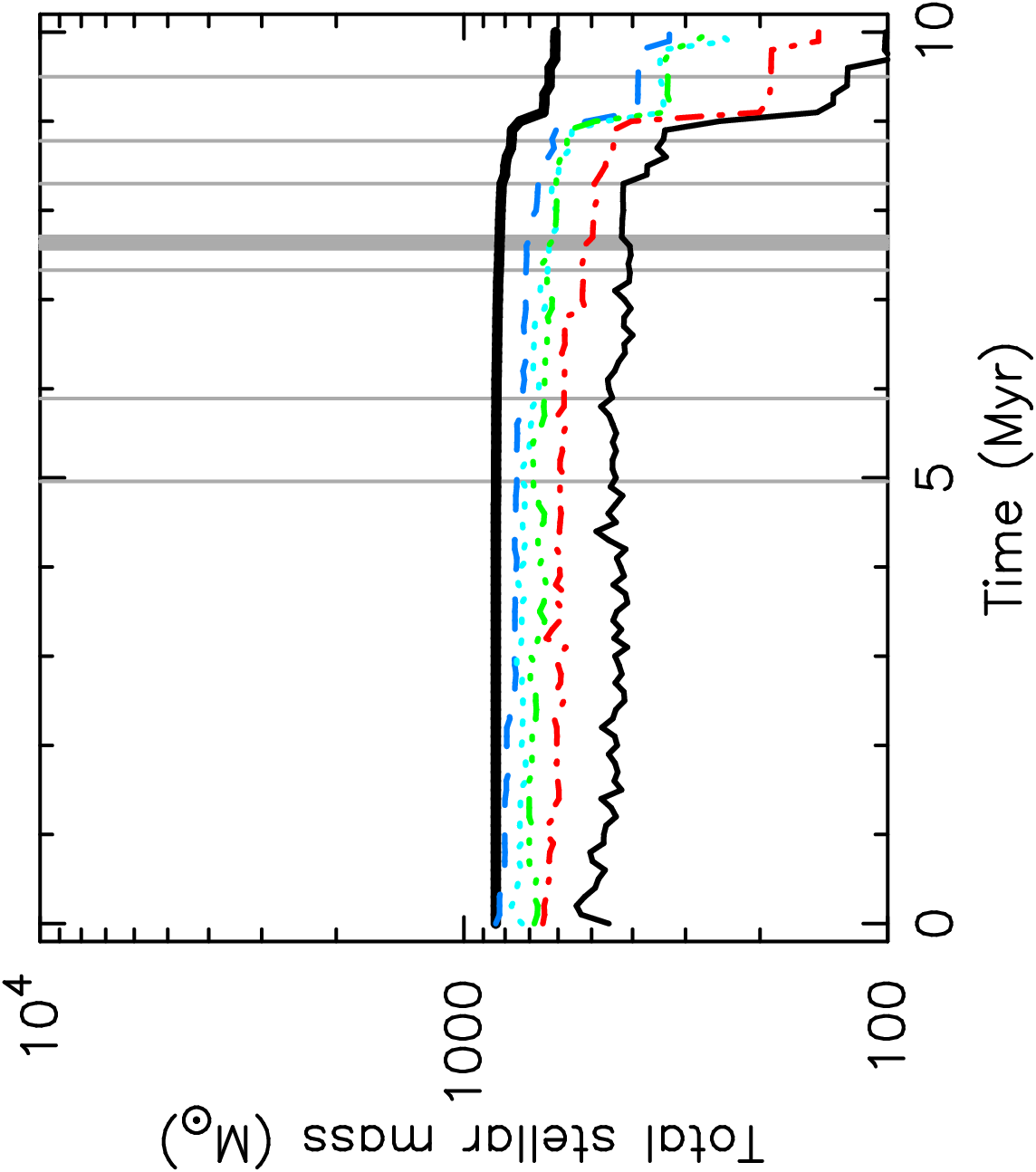}}}
    
\caption[bf]{ The evolution of total stellar mass in three of our dual-feedback run simulations in which feedback from photoionising radiation and stellar winds is implemented in the SPH runs. The thick black lines show the total mass, which decreases due to mass loss from the stars from their stellar evolution. The vertical lines indicate the time at which the first stage of post-main sequence evolution for each star that does leave the main sequence occurs. The thin lines show the evolution of the bound star mass for the versions of the simulations with different velocity scalings. The thin solid black lines are the original (sink particle) velocities, dashed dark blue lines are for $\alpha_{\rm vir} = 0.01$, dotted cyan lines are for $\alpha_{\rm vir} = 0.3$, dot-dot-dot-dashed green lines are for $\alpha_{\rm vir} = 0.5$ and dot-dashed red lines are for $\alpha_{\rm vir} = 1.5$.}  
\label{B_mass}
  \end{center}
\end{figure*}

In contrast, the runs in which the stars formed under the influence of feedback produced a more `normal', or field-like IMF, with fewer massive stars compared to the runs without feedback \citep{Dale12b,Dale14}. The overall total initial stellar mass is therefore lower in the simulations with feedback (Fig.~\ref{B_mass}). However, the lower numbers of massive stars also means that fewer stars leave the main sequence during the $N$-body integration (note the lower numbers of vertical grey lines compared to Fig.~\ref{N_mass}), and the amount of mass lost from the stars is lower.

In both Fig.~\ref{N_mass} and Fig.~\ref{B_mass} the evolution of the total mass (the thicker black lines) is the same irrespective of the initial virial ratio of the stars. However, the evolution of the \emph{bound} mass fraction depends on the initial virial ratio and the subsequent dynamical evolution, as well as the internal stellar evolution of the stars. For each simulation we show the evolution of the mass of bound stars (defined for an individual star as the total energy -- kinetic plus potential -- of the star being positive), with lines coloured according to the initial virial ratio of the stars. The thin black lines are the simulations in which the stars inherit their velocities directly from the sink particles at the end of the SPH simulations. The dark blue, cyan, green and red lines are the simulations in which the velocities of the stars are forced to virial ratios of $\alpha_{\rm vir} = 0.01, 0.3, 0.5$~and~$1.5$, respectively.

  As would be expected, the fraction of bound mass is lowest in the simulations in which the velocities of the stars are forced to be supervirial ($\alpha_{\rm vir} = 1.5$, the red lines). However, there are several interesting exceptions. In Run~J with no feedback (Fig.~\ref{N_mass-a}), the subvirial simulation in which the velocities are almost zero ($\alpha_{\rm vir} = 0.01$, the dark blue line), the region forms a Trapezium-like system, the dynamical disruption of which can sometimes unbind the entre star-forming region \citep{Allison11}. In Run~UF with feedback (Fig.~\ref{B_mass-c}), which forms stars in an unbound cloud, in the simulation where the stars inherit their velocities directly from the sink particles (the thin black line), the stars' velocities are highly supervirial ($\alpha_{\rm vir} = 2.42$). In this simulation, half of the stellar mass is unbound \emph{before} any stellar or dynamical evolution has occurred, and the bound mass fraction is always lower than the other runs, where the stars' virial ratios are nowhere near as high.

\subsection{The evolution of the virial ratio}

We next show the evolution of the virial ratio, as defined in Eqn.~\ref{vir_eq}, with the individual stellar kinetic energies and stellar potential energies calculated with Eqns.~\ref{ke_eq}~and~\ref{pe_eq}, respectively. Note that these are the virial ratios of the stars only -- the gas from the SPH simulation is not included.

\begin{figure*}
  \begin{center}
    \setlength{\subfigcapskip}{10pt}
    \hspace*{-1.3cm} \subfigure[Run J]{\label{N_virrat-a}\rotatebox{270}{\includegraphics[scale=0.285]{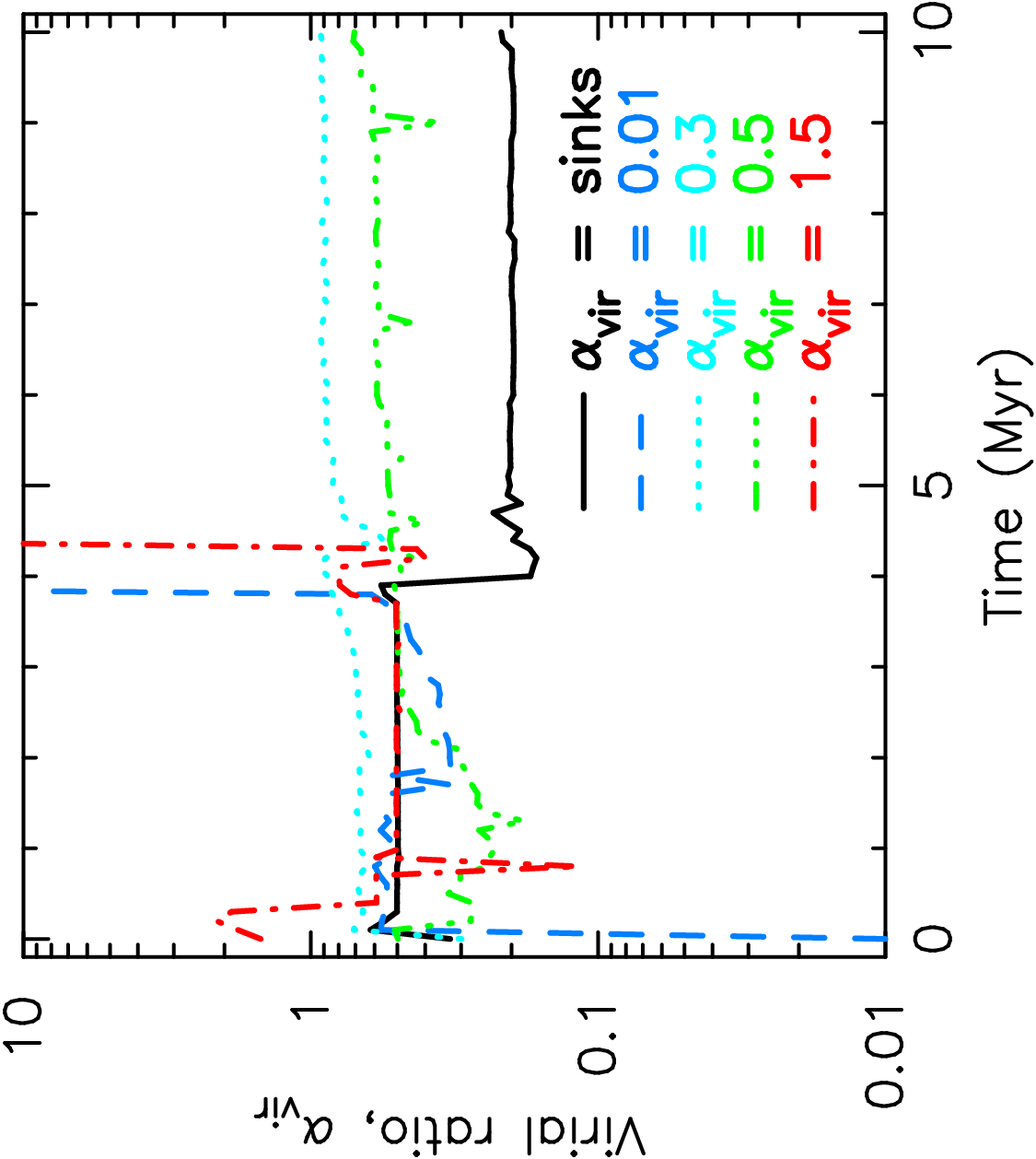}}}
    \hspace*{0.1cm} \subfigure[Run I]{\label{N_virrat-b}\rotatebox{270}{\includegraphics[scale=0.285]{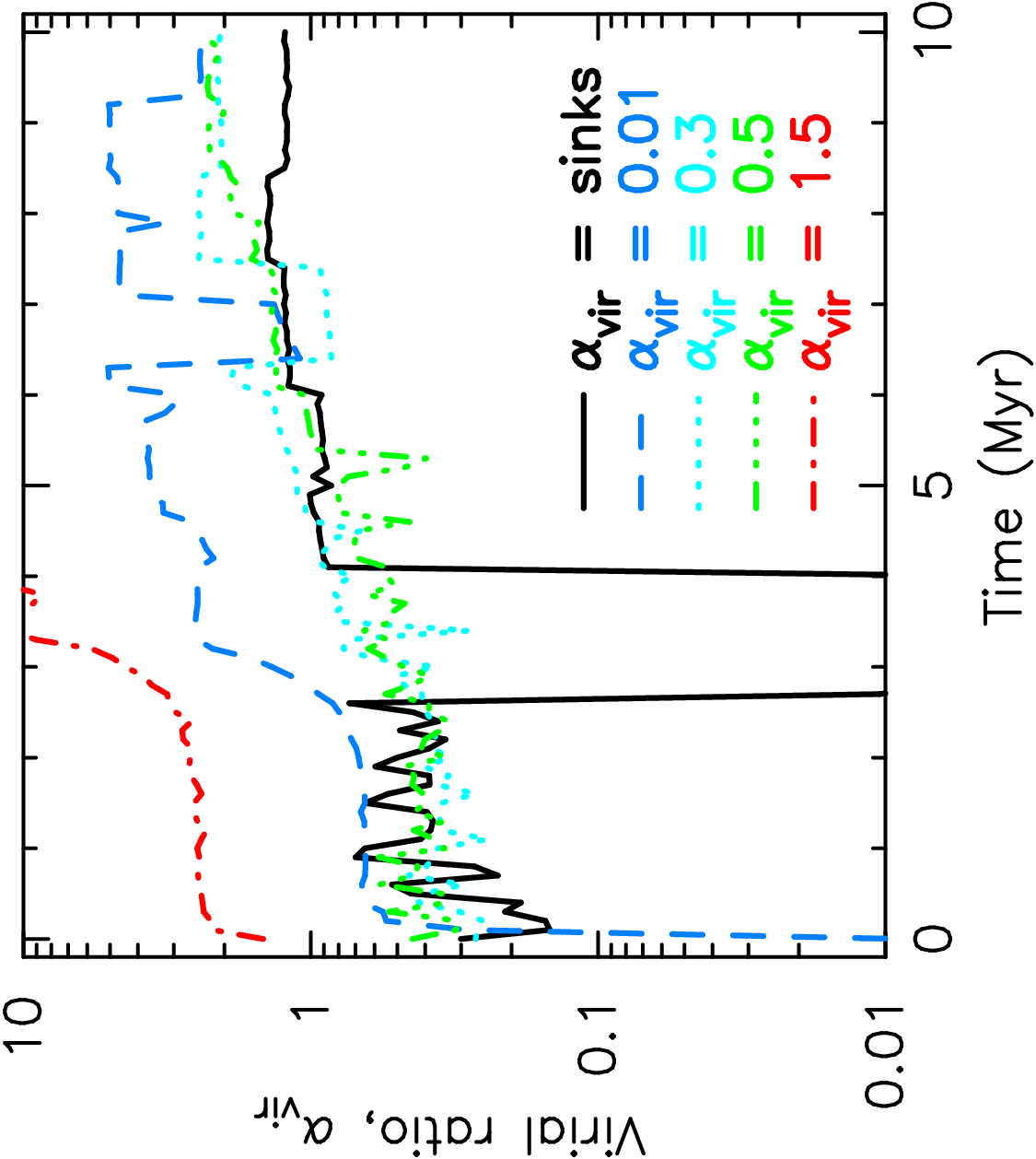}}}
    \hspace*{0.1cm} \subfigure[Run UF]{\label{N_virrat-c}\rotatebox{270}{\includegraphics[scale=0.285]{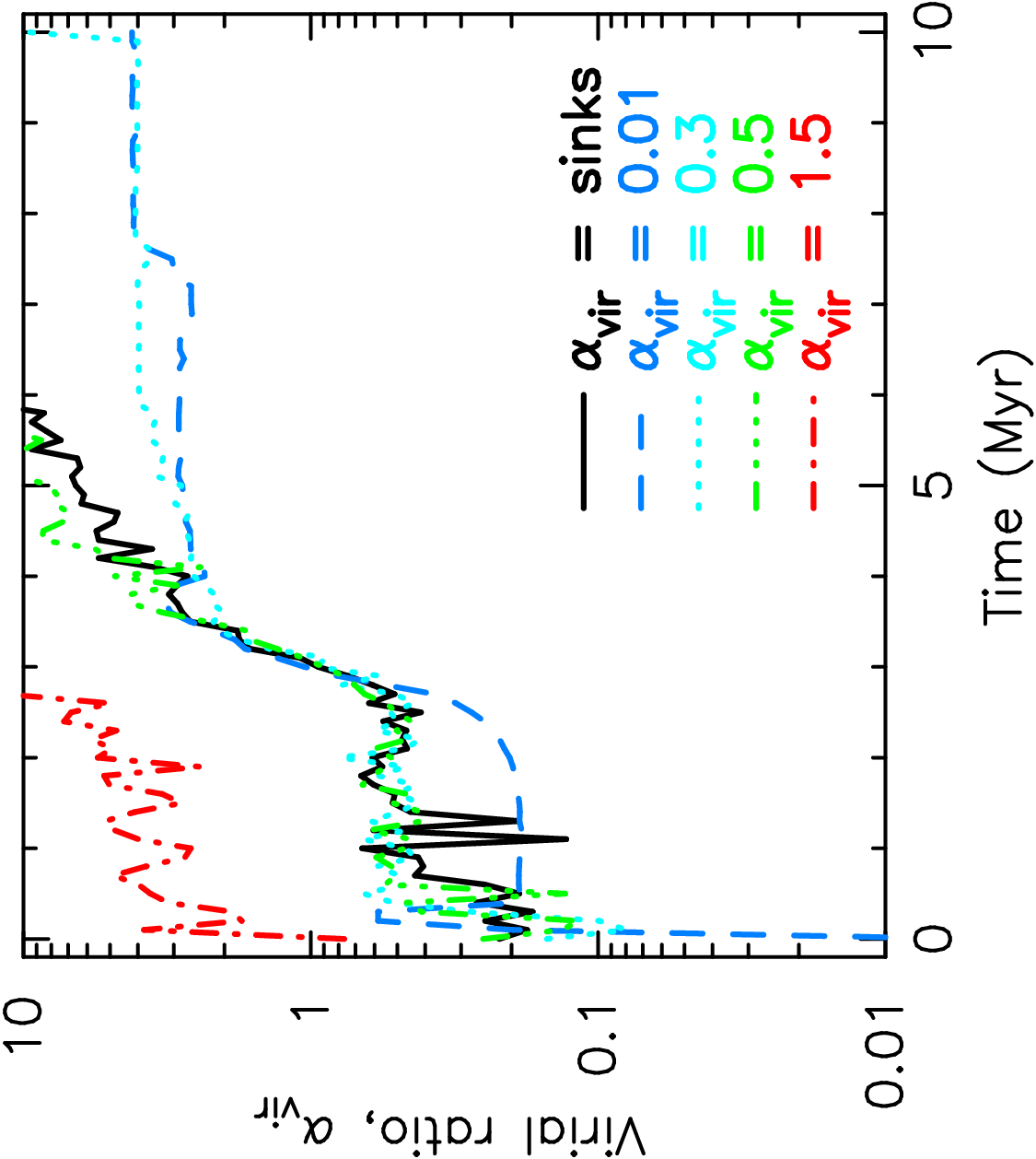}}}
    
    \caption[bf]{The evolution of the virial ratio $\alpha_{\rm vir}$ for the $N$-body simulations in which we evolve the stars from the control run simulations where no feedback is present. The different lines correspond to different initial virial ratios (i.e.\,\,we run five versions of the same initial conditions, with the only difference being the scaling of the stellar velocities). }
    
\label{N_virrat}
  \end{center}
\end{figure*}

\begin{figure*}
  \begin{center}
    \setlength{\subfigcapskip}{10pt}
    \hspace*{-1.3cm} \subfigure[Run J]{\label{B_virrat-a}\rotatebox{270}{\includegraphics[scale=0.285]{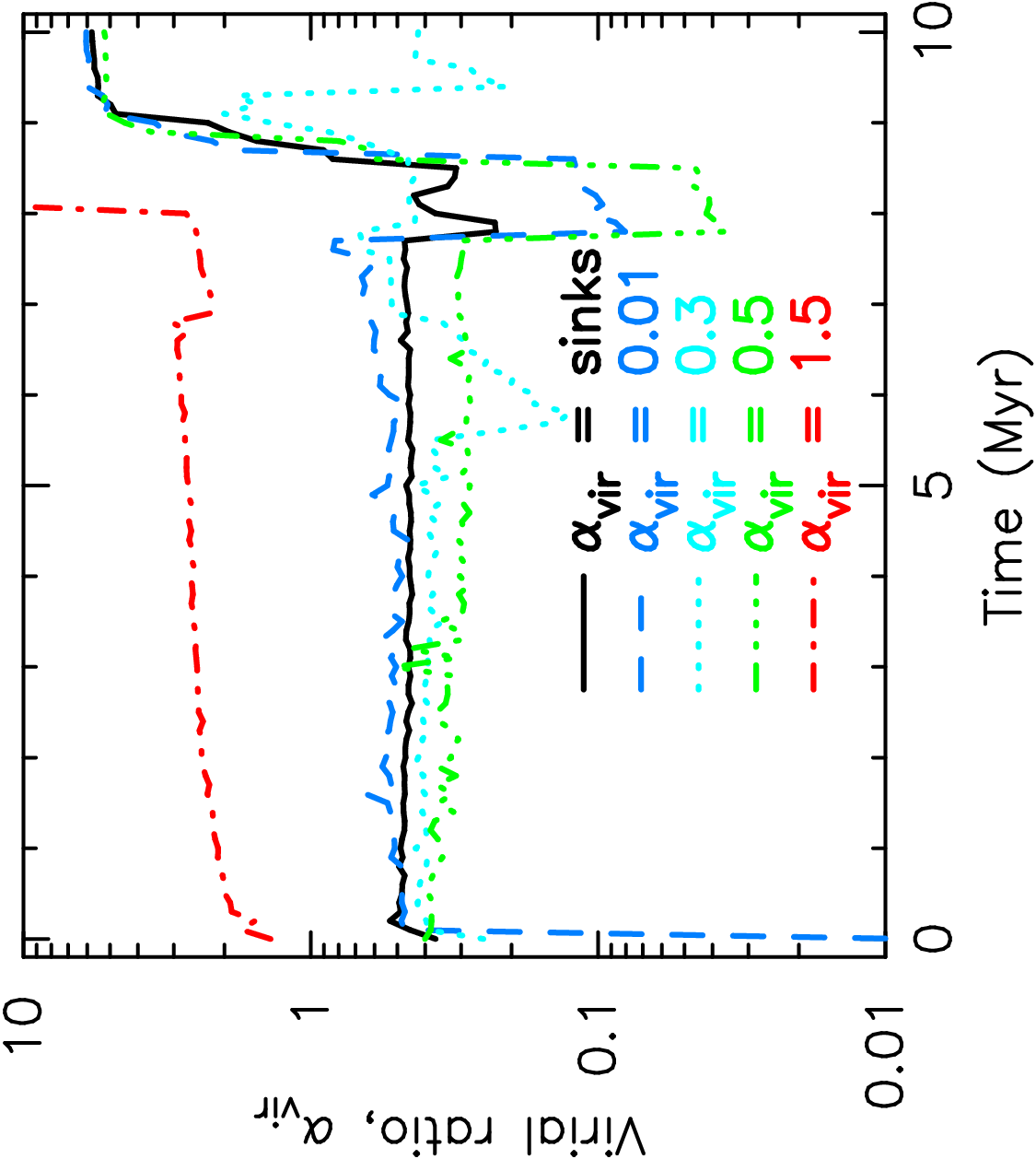}}}
    \hspace*{0.1cm} \subfigure[Run I]{\label{B_virrat-b}\rotatebox{270}{\includegraphics[scale=0.285]{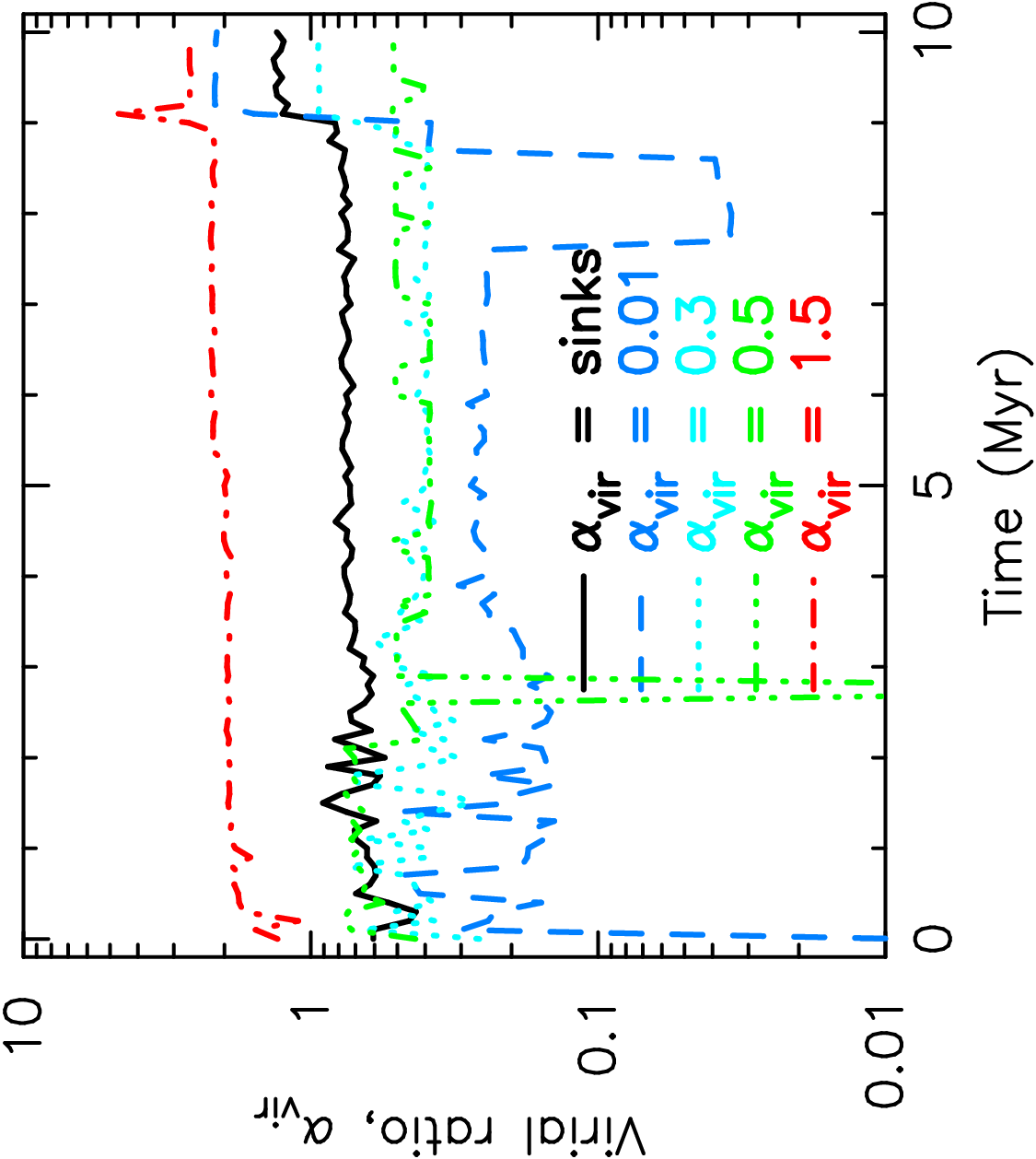}}}
    \hspace*{0.1cm} \subfigure[Run UF]{\label{B_virrat-c}\rotatebox{270}{\includegraphics[scale=0.285]{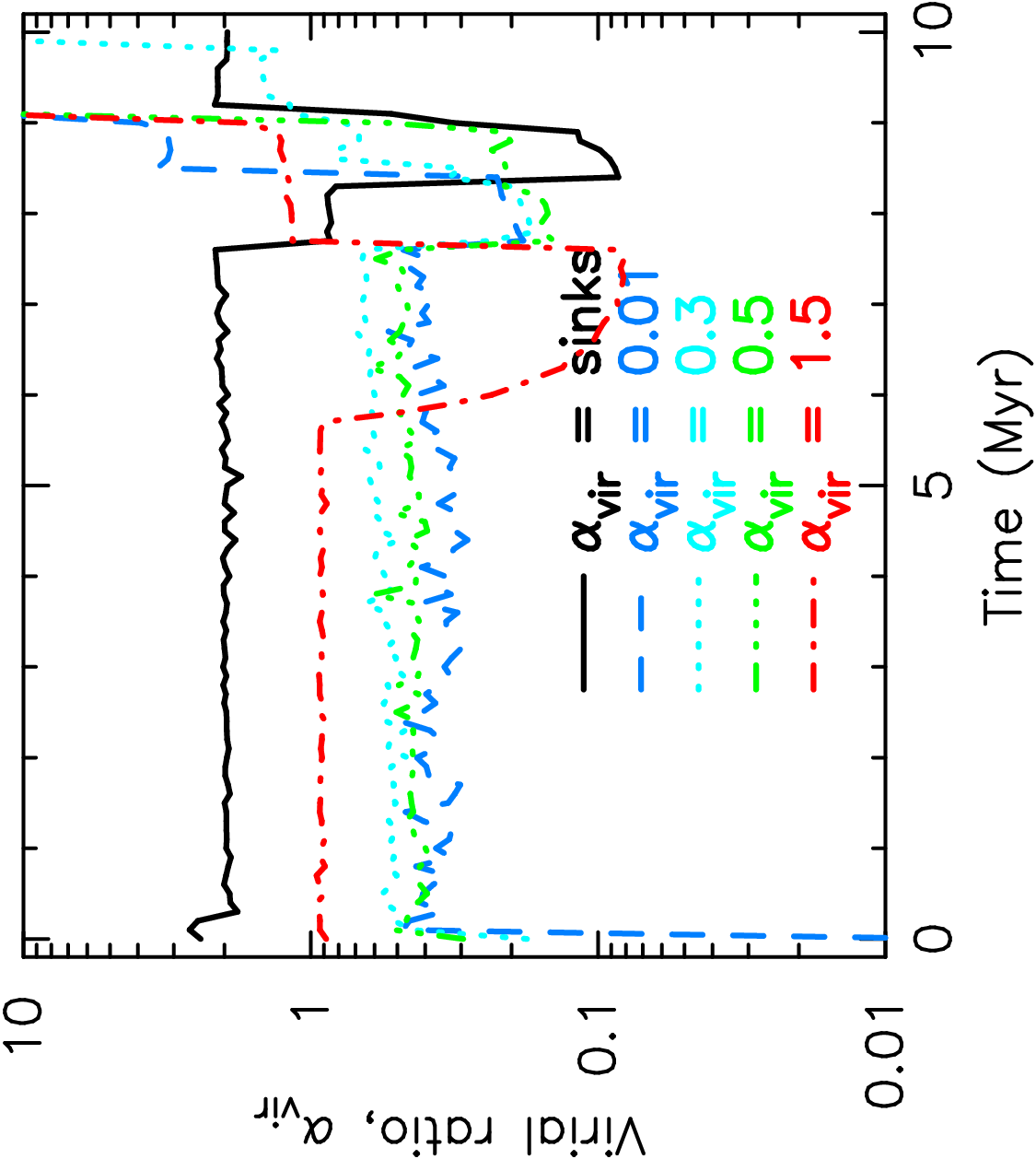}}}
    
\caption[bf]{The evolution of the virial ratio $\alpha_{\rm vir}$ for the $N$-body simulations in which we evolve the stars from the SPH simulations where feedback from photoionising radiation and stellar winds is present. The different lines correspond to different initial virial ratios (i.e.\,\,we run five versions of the same initial conditions, with the only difference being the scaling of the stellar velocities). }
\label{B_virrat}
  \end{center}
\end{figure*}

Na{\"i}vely we would expect the sudden removal of the gas -- a significant contribution to the gravitational potential energy before the $N$-body calculations -- to cause the velocity dispersion of the stars to be supervirial, in that they should be moving faster due to the potential energy from the gas. However, the virial ratio in the simulations where we take the stellar velocities directly from the sink particles in the final SPH snapshot is often close to virial equilibrium $\alpha_{\rm vir} = 0.5$, or subvirial  $\alpha_{\rm vir} < 0.5$ -- see the black lines at 0\,Myr in Figs.~\ref{N_virrat}~and~\ref{B_virrat}, and the values in Table~\ref{cluster_props}.

In the simulations in which we scale the stellar velocities to a different virial ratio, the virial ratio rapidly evolves to values around $\alpha_{\rm vir} = 0.5$ within the first 0.5\,Myr (e.g. the dark blue dashed, cyan dotted, and green dashed-dot-dot lines), including the simulation in which we scale the velocities to a very low initial virial ratio of $\alpha_{\rm vir} = 0.01$ (the blue dashed line). The only exception to this is the simulation in  which the velocities are scaled to be highly supervirial ($\alpha_{\rm vir} = 1.5$, the red dot-dashed lines). In these simulations the virial ratio generally remains supervirial.

There are various points in some simulations where sudden mass-loss from the stars causes a sudden increase in the virial ratio (and these increases in virial ratio correspond to stars leaving the main sequence, as indicated by the vertical lines in Figs.~\ref{N_mass}~and~\ref{B_mass}). In Fig.~\ref{N_virrat} we show the evolution of the virial ratio for the $N$-body evolution of the simulations that formed stars with no feedback. As these simulations have a top-heavy IMF, mass loss from the stars due to stellar evolution is both more drastic, and happens earlier. In Runs J and I (Figs.~\ref{N_virrat-a}~and~\ref{N_virrat-b}), significant mass-loss occurs after 4\,Myr, and there are noticable increases in the virial ratios at these times. In Run UF (Fig.~\ref{N_virrat-c}), the mass-loss is such a high fraction of the potential energy that the virial ratios in all simulations become supervirial.

In the simulations in which the sink particles form with feedback, the mass functions are more `normal' and subsequent stellar evolution does not have as high an impact on the evolution of the star-forming regions, as there is less mass-loss from the stars. This is seen in the evolution of the virial ratio in these simulations (Fig.~\ref{B_virrat}), where the virial ratios only systematically increase towards the end of the $N$-body evolution, at 8 -- 9\,Myr. With the exception of Run UF (Fig.~\ref{B_virrat-c}), the virial ratios of the simulations where the sink velocities are not scaled are very similar to the simulations in which the sink velocities are scaled to (sub)virial velocities (compare the black lines with the green and blue lines). 

In both Fig.~\ref{N_virrat}~and~\ref{B_virrat} there are instances where the virial ratio suddenly decreases to extremely small values, which occurs due to the formation of an extremely close binary system. Close binary systems ($<1$\,au) are known to dominate the potential energy of a star-forming region, and observational studies of star-forming regions attempt to correct for the orbital motion of close binary systems \citep[e.g.][]{Gieles10,Cottaar12b,Cottaar14a}. We have not applied a correction because the close binary systems that form here eventually merge, and the virial ratio then returns to its previous value(s). 

In each simulation there is a population of wide ($>100$\,au) binary stars, which typically comprise around 10\,per cent of the total number of stellar systems in the simulation. Aside from the occasional formation of a close binary which can affect the overall virial ratio (as detailed above), the population of wide binaries does not affect the dynamical evolution of the star-forming regions, and does not bias the determination of the $\mathcal{Q}$-parameter or the $\Lambda_{\rm MSR}$ mass segregation ratio.

\subsection{Evolution of the half-mass radius}

\begin{figure*}
  \begin{center}
    \setlength{\subfigcapskip}{10pt}
    \hspace*{-1.3cm} \subfigure[Run J]{\label{N_rhm-a}\rotatebox{270}{\includegraphics[scale=0.285]{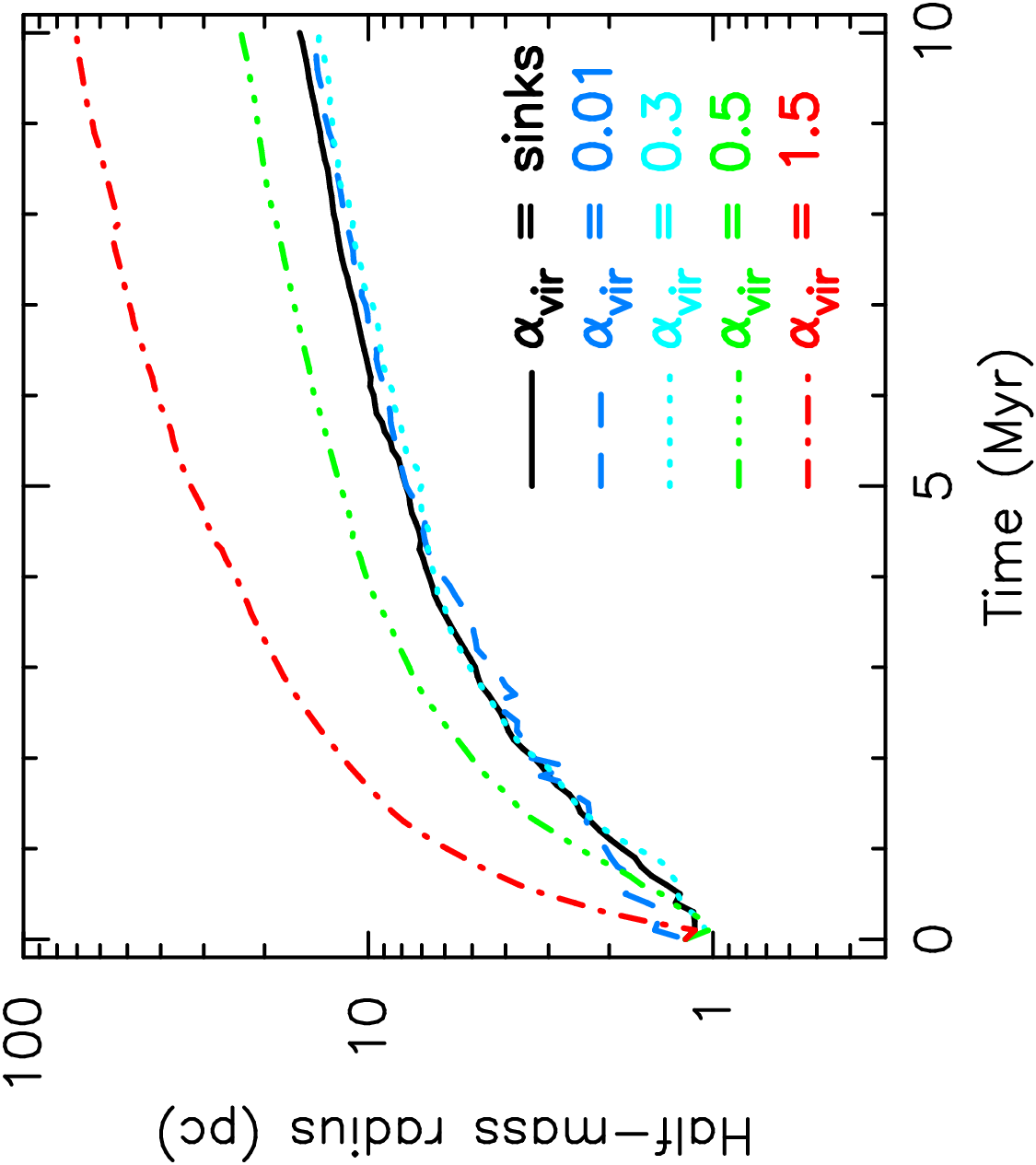}}}
    \hspace*{0.1cm} \subfigure[Run I]{\label{N_rhm-b}\rotatebox{270}{\includegraphics[scale=0.285]{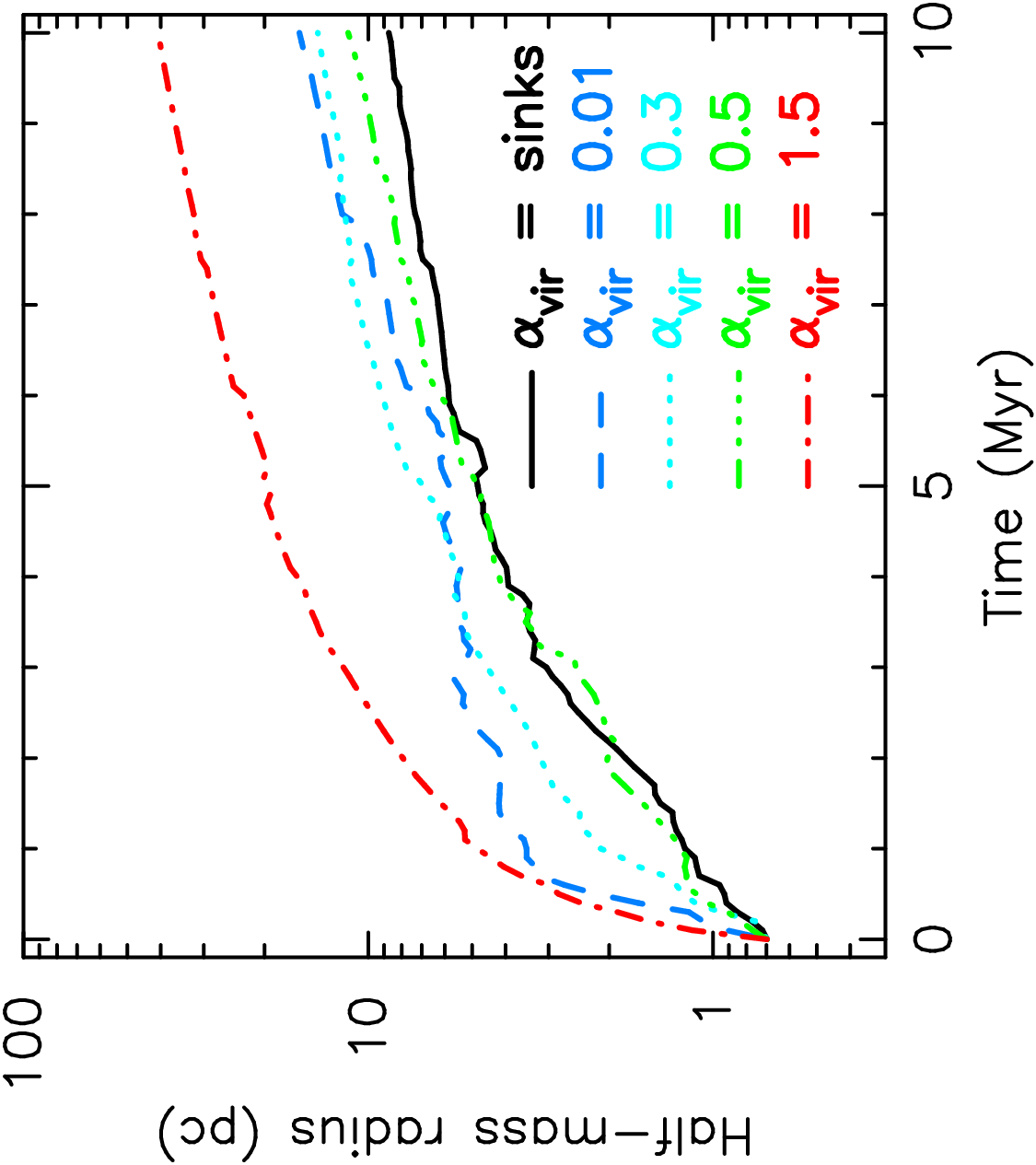}}}
    \hspace*{0.1cm} \subfigure[Run UF]{\label{N_rhm-c}\rotatebox{270}{\includegraphics[scale=0.285]{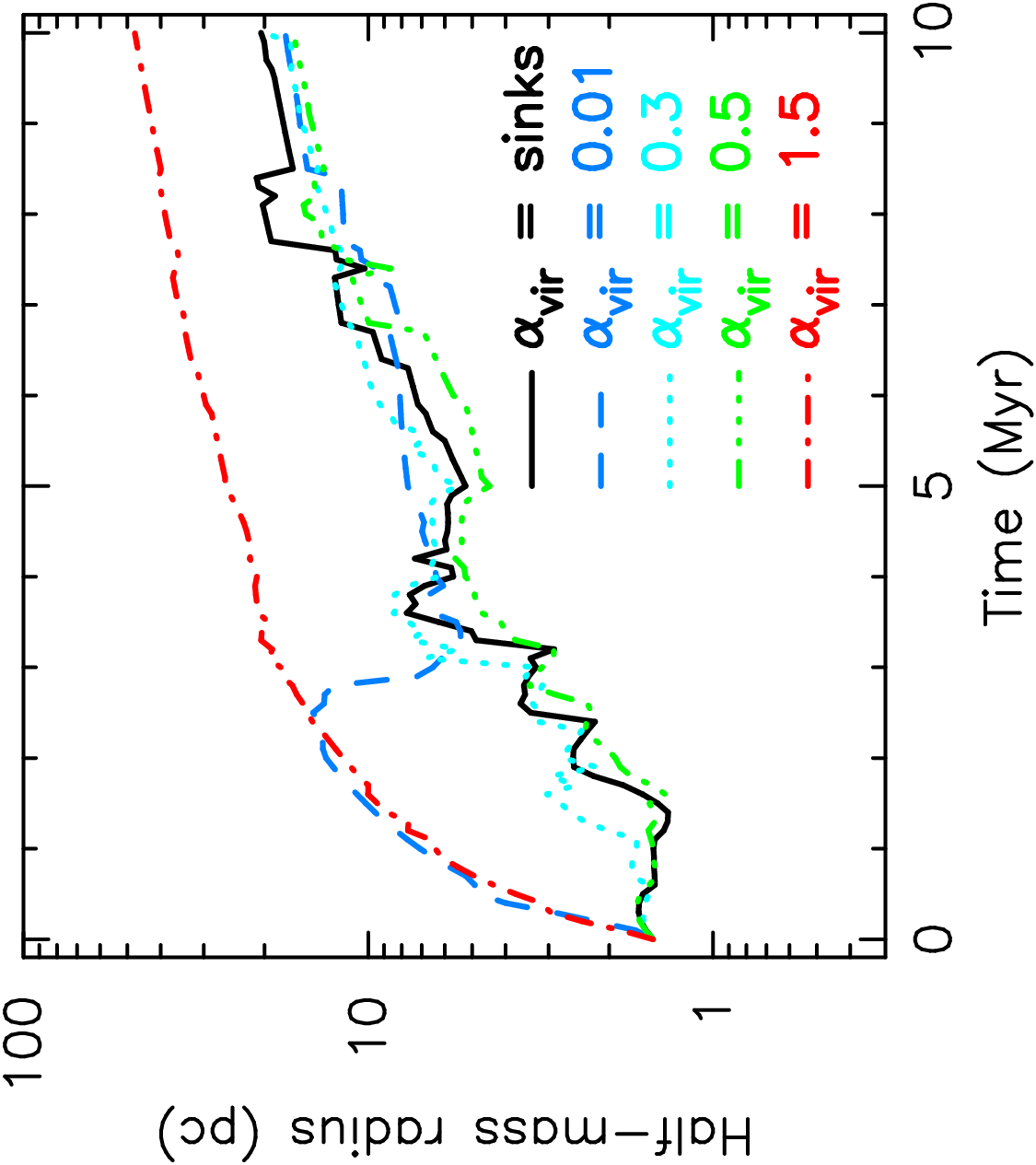}}}
    
\caption[bf]{The evolution of the half-mass radius with time for $N$-body simulations of stars that formed in the SPH control runs with no feedback. The different lines correspond to different initial virial ratios (i.e.\,\,we run five versions of the same initial conditions, with the only difference being the scaling of the stellar velocities).  }
\label{N_rhm}
  \end{center}
\end{figure*}

\begin{figure*}
  \begin{center}
    \setlength{\subfigcapskip}{10pt}
    \hspace*{-1.3cm} \subfigure[Run J]{\label{B_rhm-a}\rotatebox{270}{\includegraphics[scale=0.285]{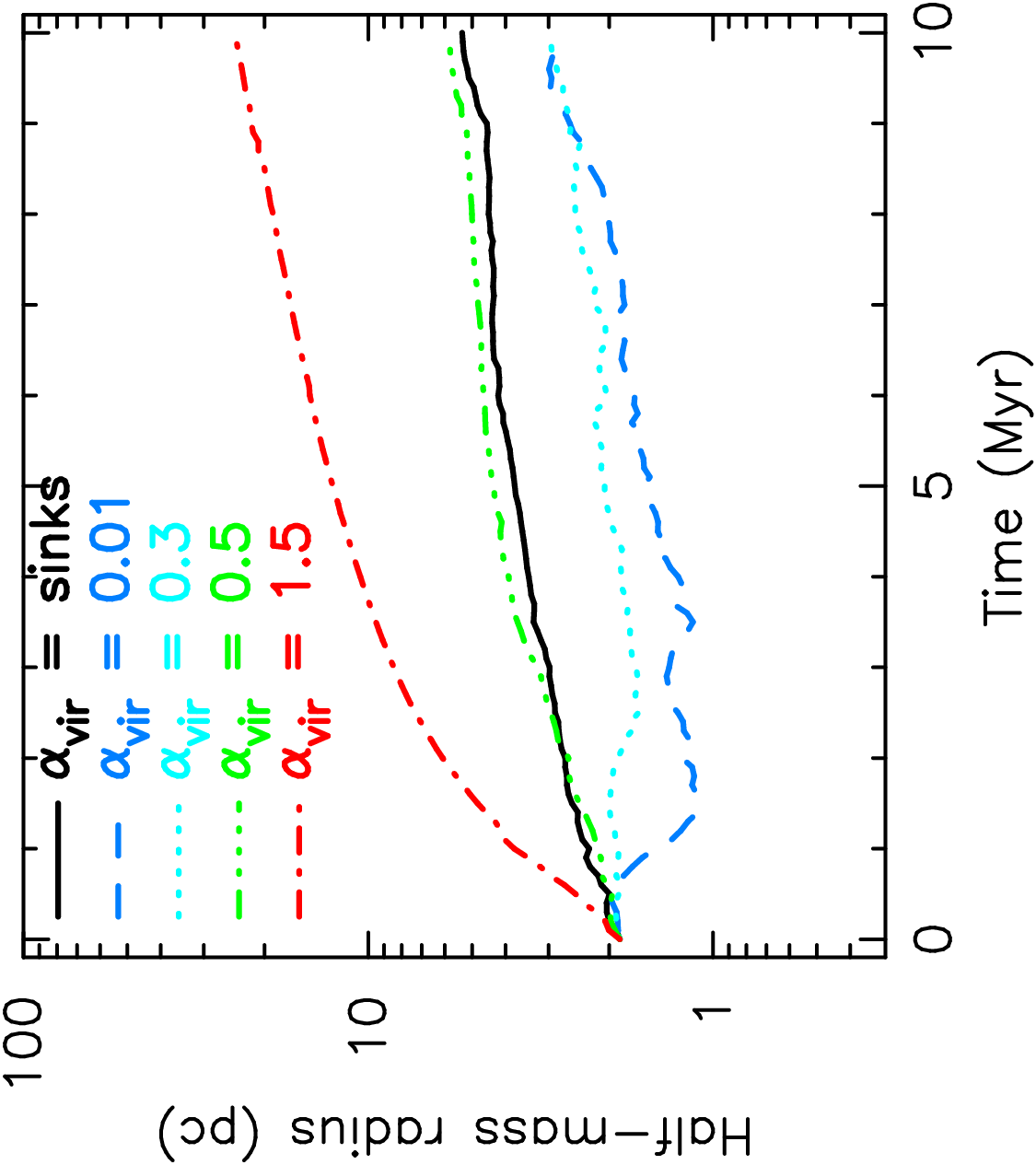}}}
    \hspace*{0.1cm} \subfigure[Run I]{\label{B_rhm-b}\rotatebox{270}{\includegraphics[scale=0.285]{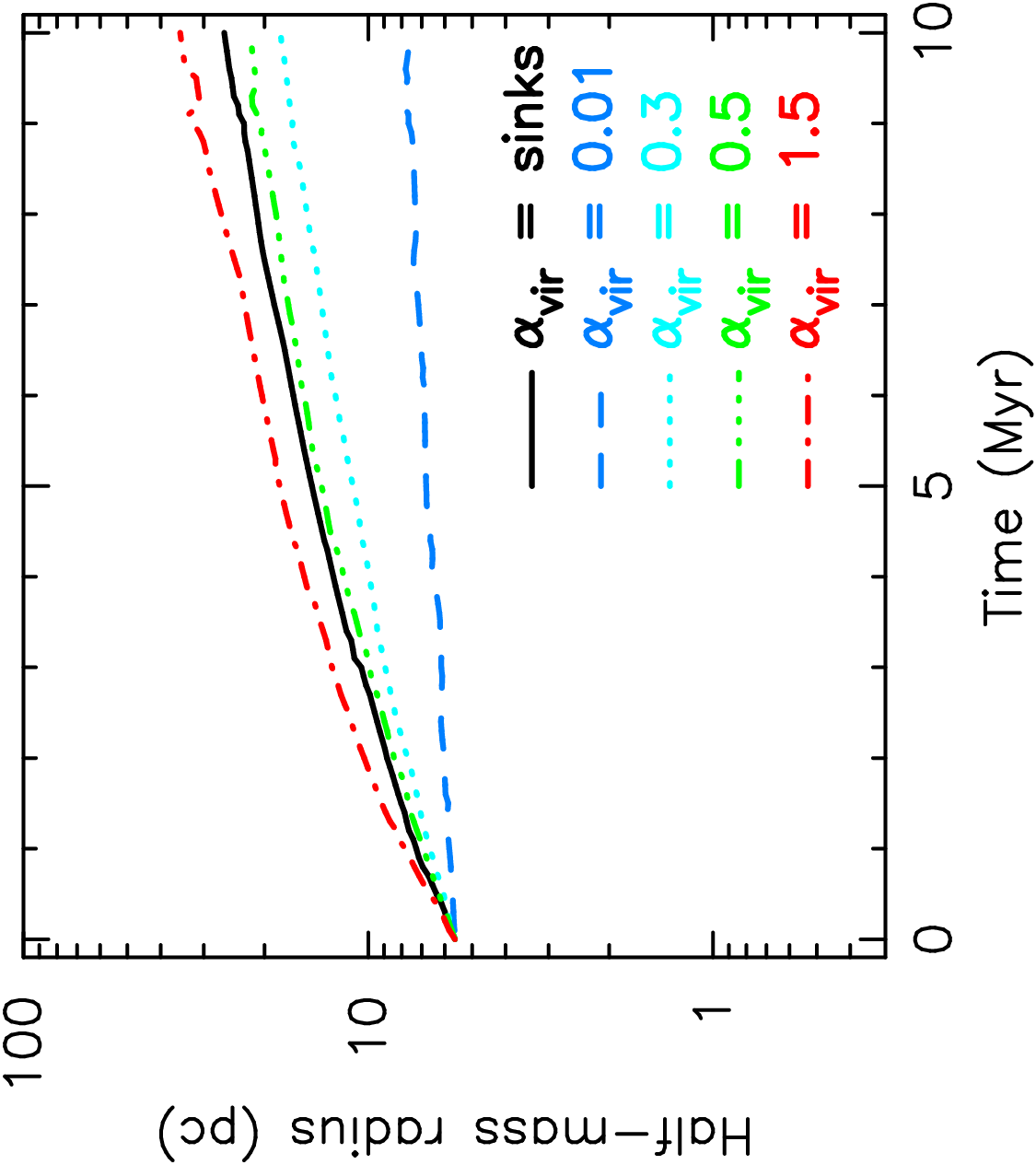}}}
    \hspace*{0.1cm} \subfigure[Run UF]{\label{B_rhm-c}\rotatebox{270}{\includegraphics[scale=0.285]{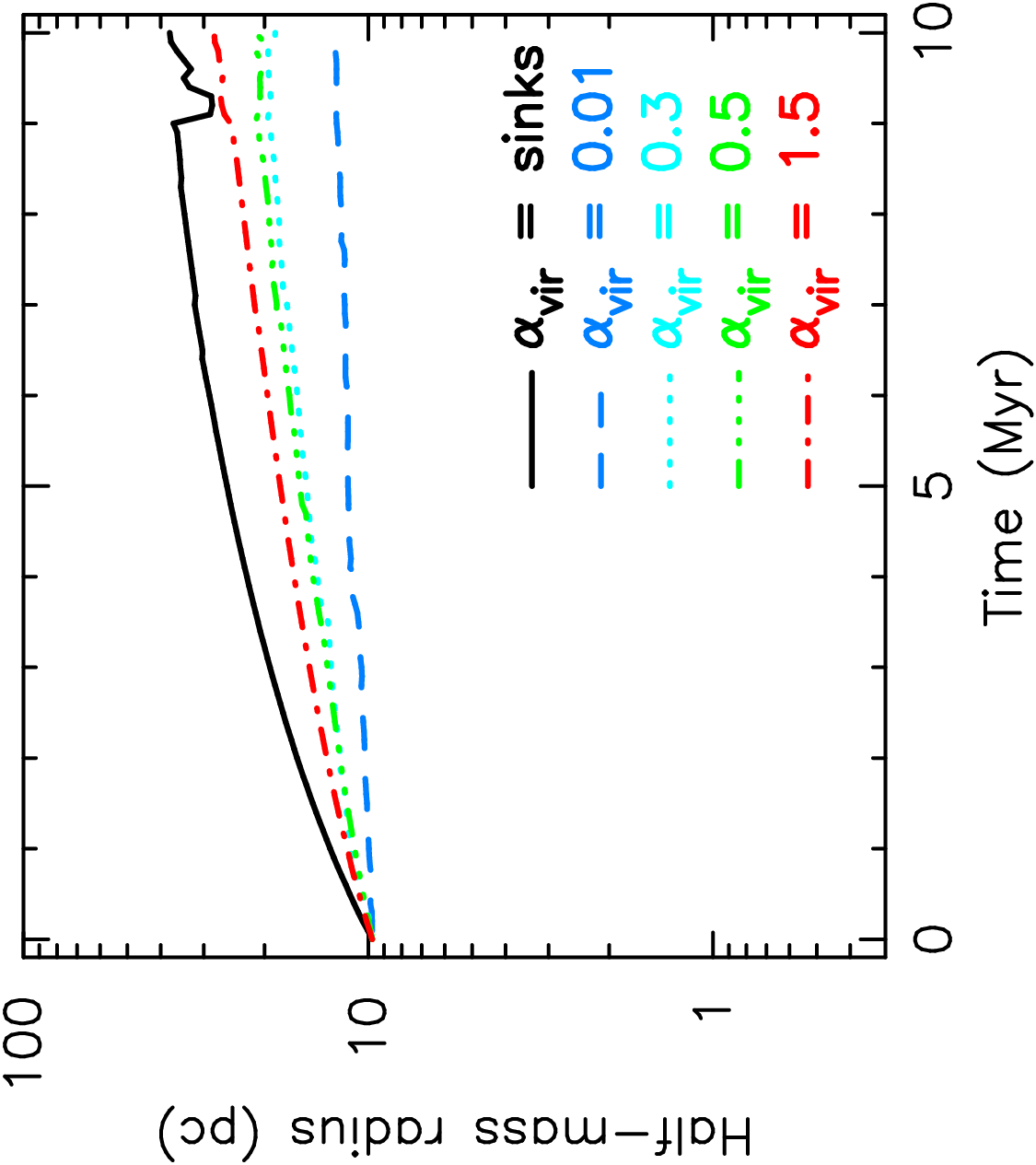}}}
    
\caption[bf]{The evolution of the half-mass radius with time for $N$-body simulations of stars that formed in the SPH simulations with feedback from photoionising radiation and stellar winds present. The different lines correspond to different initial virial ratios (i.e.\,\,we run five versions of the same initial conditions, with the only difference being the scaling of the stellar velocities). }
\label{B_rhm}
  \end{center}
\end{figure*}

We now show the evolution of the half-mass radius, as defined in Section~\ref{method}, for the $N$-body evolution of simulations that formed stars without feedback (Fig.~\ref{N_rhm}) and the simulations that formed stars with feedback (Fig.~\ref{B_rhm}).

Because the half-mass radius is measured from the positions of the stars, the initial half-mass radii in all simulations are identical, but diverge due to the different velocity distributions.

In the simulations where the stars form with no feedback present (Fig.~\ref{N_rhm}), the different versions of the $N$-body evolution where the stars are scaled to different virial ratios are broadly similar between the different simulations. The simulations where the stars are scaled to supervirial velocities (the red dot-dashed lines) evolve to larger half-mass radii as the star-forming regions rapidly expand.

For the simulations where the stars retain their velocities from the end of the SPH calculations (the solid black lines) and the simulations where we scale the velocities to be (sub)virial, the half-mass radii evolution overlaps throughout. This suggests that the evolution of the regions with the unscaled velocity simulations is more akin to a region in virial equilibrium, rather than being supervirial due to the expulsion of the gas leftover from star formation.

This effect is more obvious in the $N$-body evolution of the simulations where the stars formed under the influence of feedback (Fig.~\ref{B_rhm}). These simulations have lower density to begin with (the initial half-mass radii are several pc, or even as high as 10\,pc) compared to the simulations that formed without feedback, that have initial half-mass radii of $\sim$1\,pc or less. In these simulations the evolution of the half-mass radius is entirely dependent on the initial virial ratio of the stars. The highly supervirial regions (the red dot-dashed lines) have the highest half-mass radii, and the most subvirial regions (the dark blue dashed lines) have the smallest half-mass radii. The regions that have slightly subvirial, or virial velocities have an evolution somewhere in the middle of these extremes. The simulations in which the stars just inherit the sink particle velocities evolve according to the virial ratio of those sink particles. For Runs J and I (Figs.~\ref{B_rhm-a}~and~\ref{B_rhm-b}) the initial virial ratios are around 0.5, and so these simulations evolve in a similar manner to the (sub)virial simulations. For Run UF (Fig.~\ref{B_rhm-c}) the virial ratio of the sink particles is supervirial $\alpha_{\rm vir} = 2.42$, even more supervirial than the simulation where we force the velocities to be supervirial ($\alpha_{\rm vir} = 1.5$), and so this simulation evolves to have the highest half-mass radius. This high virial ratio causes the rapid expansion of the star-forming region, which results in the half-mass radius being higher than in the other versions of this simulation. 

\subsection{Evolution of median surface density}

The evolution of the half-mass radius is a reasonable indicator of the amount of dynamical relaxation that is taking place in the star-forming regions, but can be unreliable if the region  is not centrally concentrated (the definition of mass enclosed within  a radius from a central point is  ambiguous if the distribution is substructured), and it is also difficult to make a direct comparison with observations. For these reasons, we also show the evolution of two dimensional measures that can be directly compared with observational data \citep[e.g.][]{Hetem15,Belen16,Gonzalez17,Buckner19,Hetem19,Dib18,Dib19,Rodriguez23,Coenda25}.

\begin{figure*}
  \begin{center}
    \setlength{\subfigcapskip}{10pt}
    \hspace*{-1.5cm} \subfigure[Run J]{\label{N_Sig-a}\rotatebox{270}{\includegraphics[scale=0.2875]{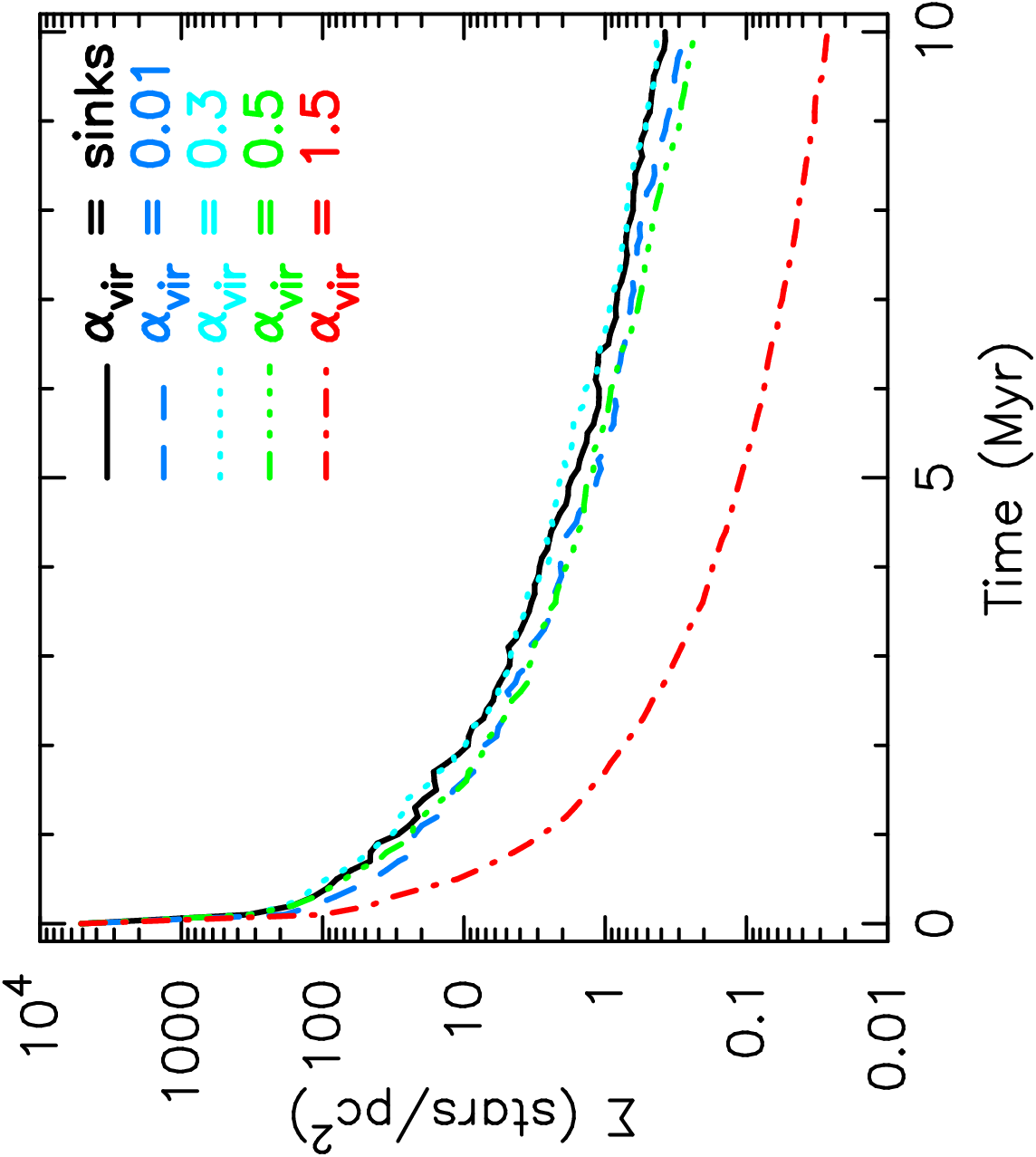}}}
    \hspace*{-0cm} \subfigure[Run I]{\label{N_Sig-b}\rotatebox{270}{\includegraphics[scale=0.2875]{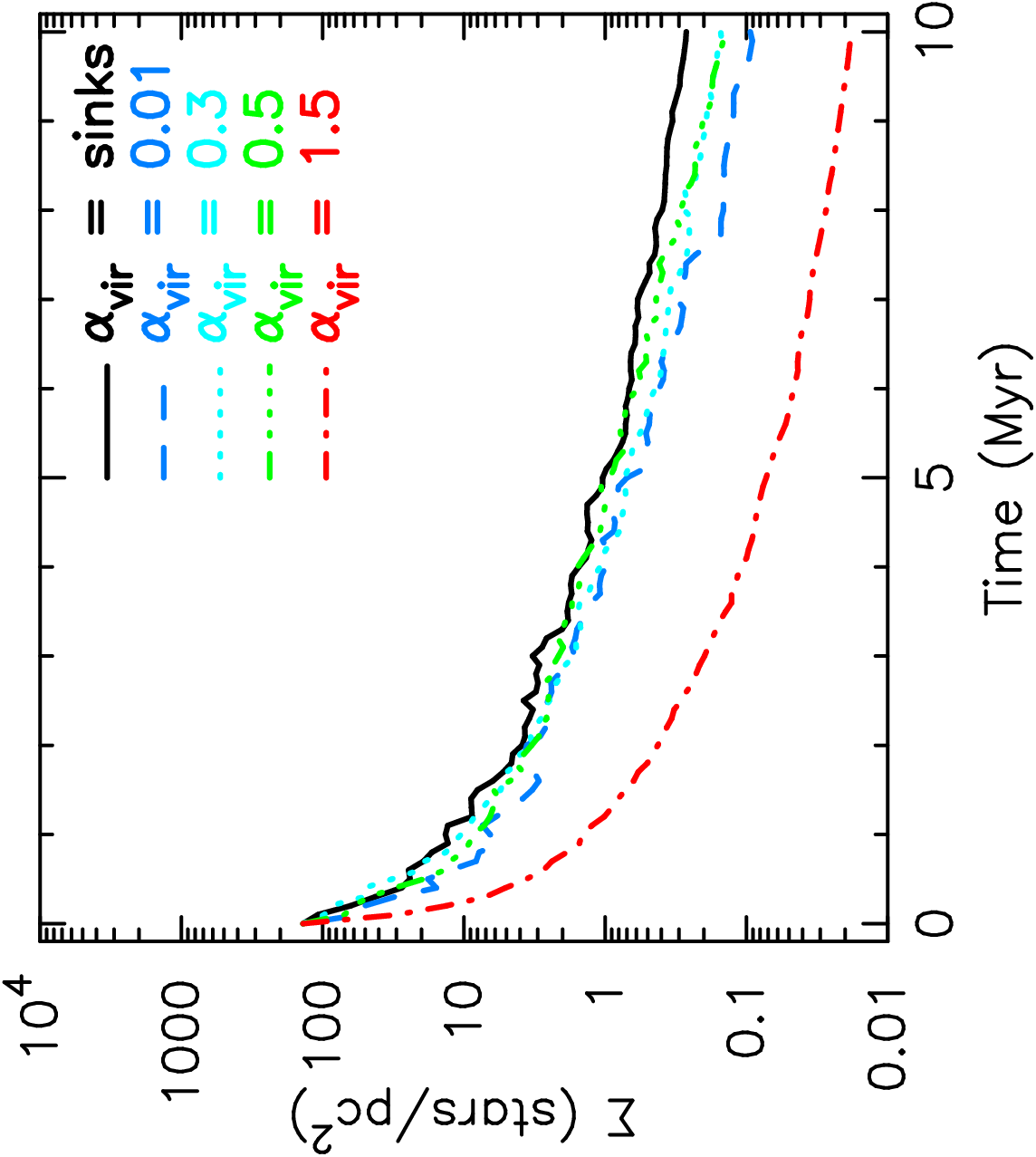}}}
    \hspace*{-0cm} \subfigure[Run UF]{\label{N_Sig-c}\rotatebox{270}{\includegraphics[scale=0.2875]{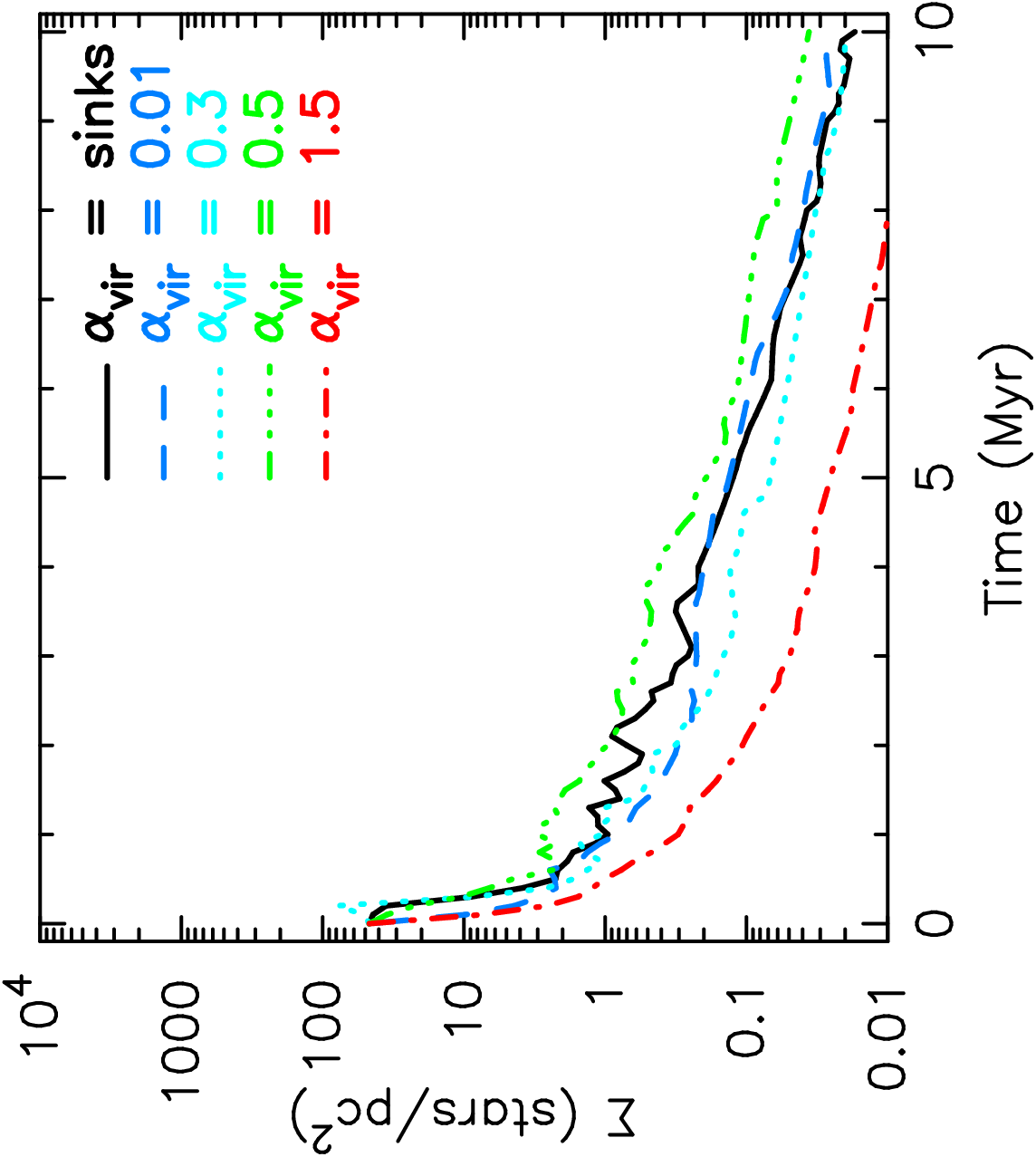}}}


\caption[bf]{Evolution of the median stellar surface density with time in $N$-body simulations where the stars form in a control run SPH simulation without feedback present. The different lines correspond to different initial virial ratios (i.e.\,\,we run five versions of the same initial conditions, with the only difference being the scaling of the stellar velocities).}
\label{N_Sig}
  \end{center}
\end{figure*}

\begin{figure*}
  \begin{center}
    \setlength{\subfigcapskip}{10pt}
    \hspace*{-1.5cm} \subfigure[Run J]{\label{B_Sig-a}\rotatebox{270}{\includegraphics[scale=0.2875]{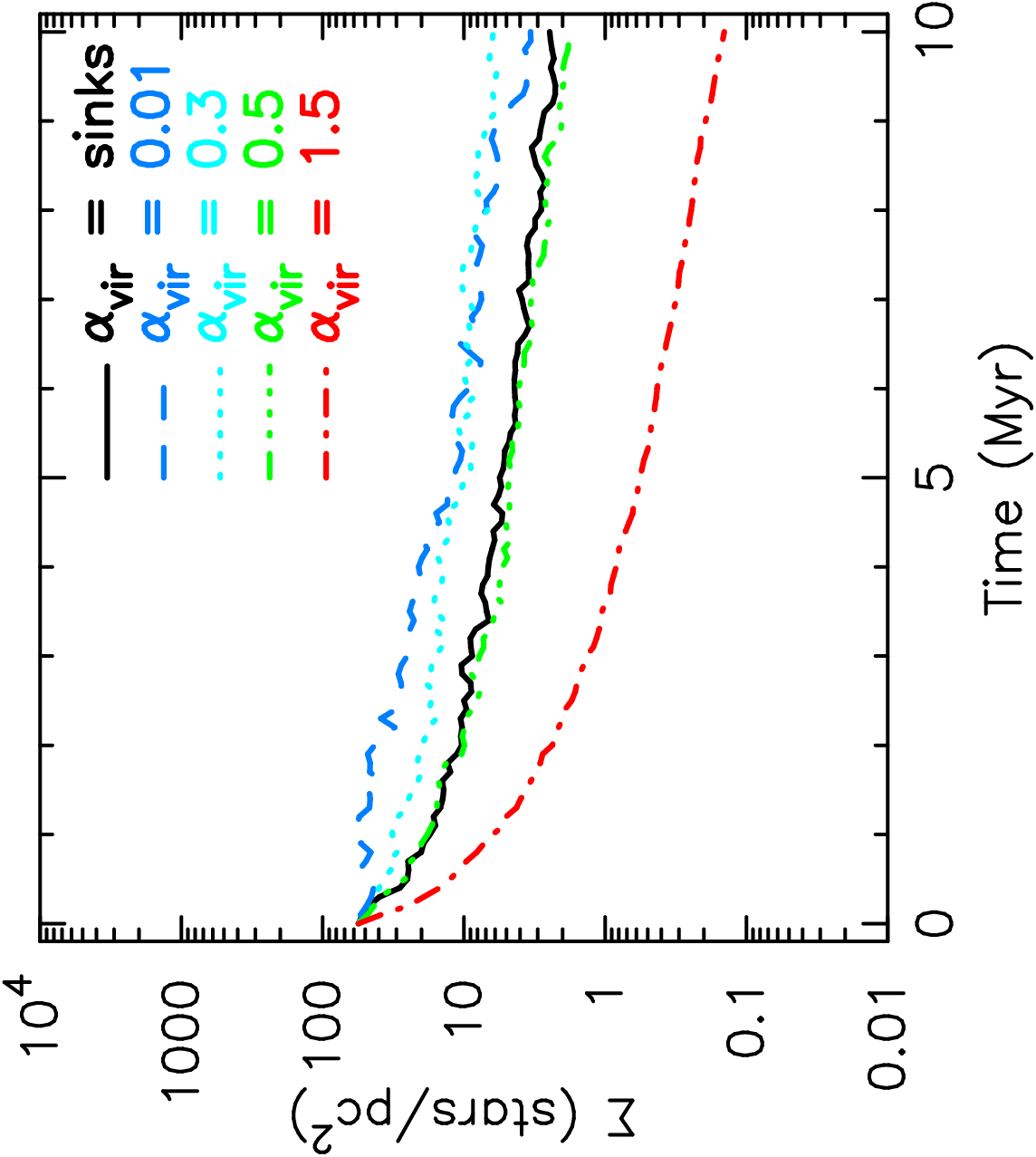}}}
    \hspace*{-0cm} \subfigure[Run I]{\label{B_Sig-b}\rotatebox{270}{\includegraphics[scale=0.2875]{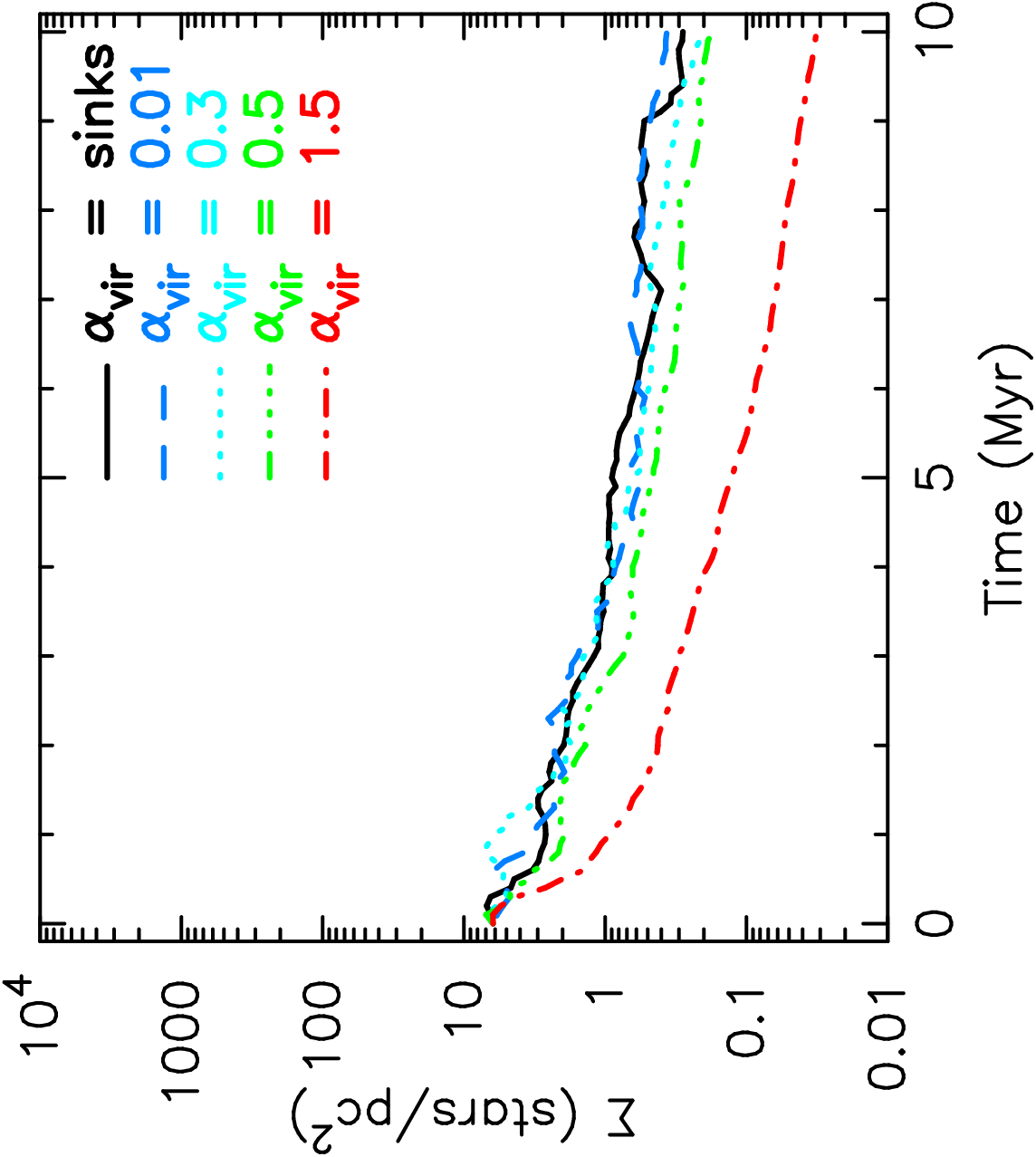}}}
    \hspace*{-0cm} \subfigure[Run UF]{\label{B_Sig-c}\rotatebox{270}{\includegraphics[scale=0.2875]{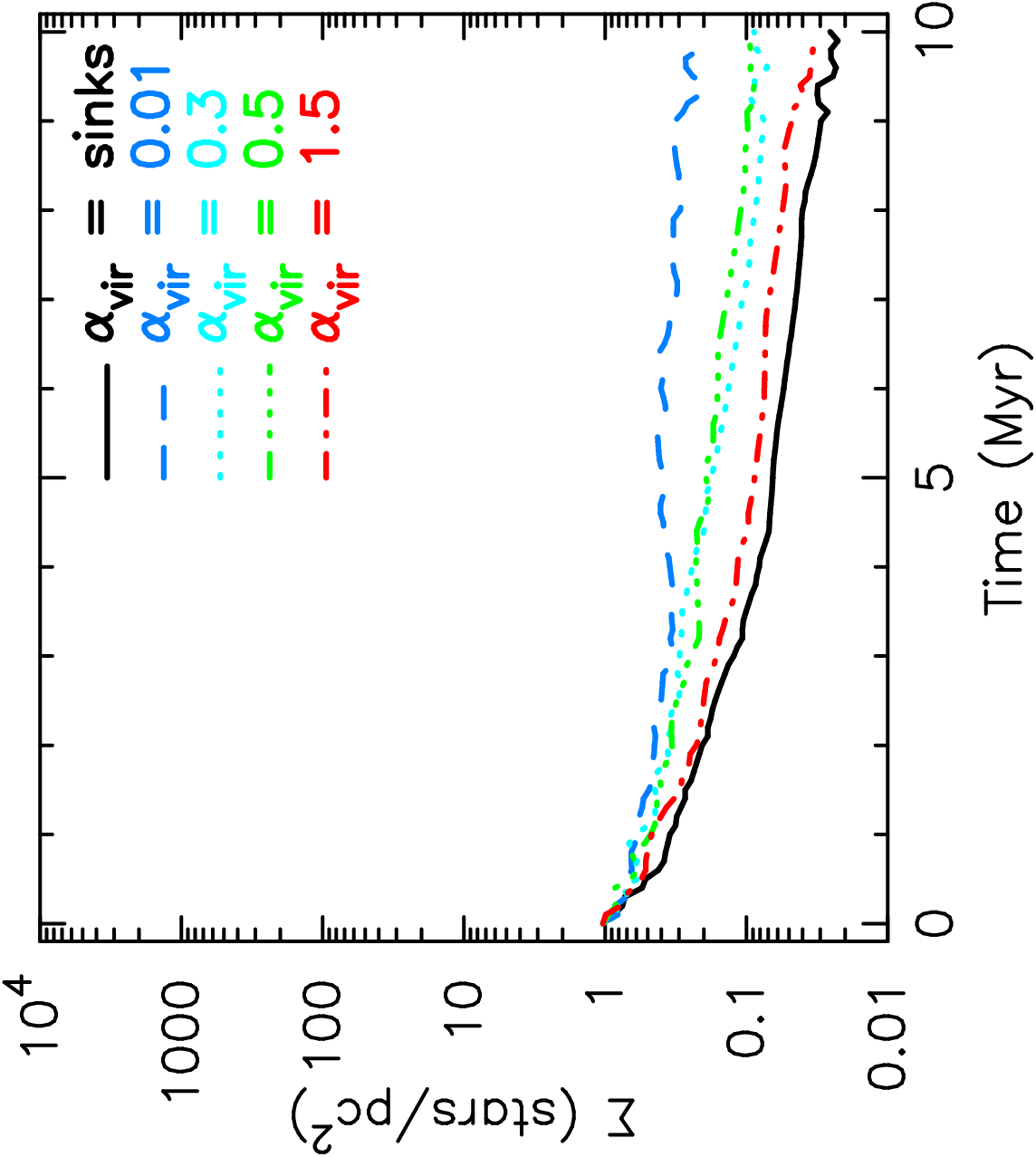}}}


\caption[bf]{The evolution of the median stellar surface density with time for $N$-body simulations of stars that formed in the SPH simulations with feedback from photoionising radiation and stellar winds. The different lines correspond to different initial virial ratios (i.e.\,\,we run five versions of the same initial conditions, with the only difference being the scaling of the stellar velocities). }
\label{B_Sig}
  \end{center}
\end{figure*}

The first of these two-dimensional measures is the median value of the individual local stellar surface densities, as defined by Eqn.~\ref{sig_eqn}. Despite only utilising two dimensions (we use the $x-y$ plane, to mimic the Right Ascension and Declination available to observers), the measure is fairly robust against the effects of complex geometries \citep[e.g.][]{Parker12d}.

We show the evolution with time of the median surface density for the stars that form in the SPH simulations where no feedback is present in Fig.~\ref{N_Sig}. The different lines correspond to the stars being scaled to different virial ratios. The simulations that are scaled to supervirial velocities quickly evolve to very low median surface densities (the red dot-dashed lines). However, the remaining realisations all evolve in a similar fashion, and it is difficult to distinguish between different initial virial states.

When the stars form under the influence of feedback from photoionising radiation and stellar winds (Fig.~\ref{B_Sig}), the evolution of the median surface density is qualitatively similar to the control run cases. The supervirial simulations evolve to the lowest densities. However, in Run J (Fig.~\ref{B_Sig-a}) and Run UF (Fig.~\ref{B_Sig-c}) the subvirial simulations evolve such that they have higher densities than the virialised and supervirial simulations. The realisation of the simulation that inherits the sink particle velocities (i.e. there is no scaling to a virial ratio) in Run UF follows a similar evolution to the supervirial simulation, ostensibly because the sink particles happen to have supervirial velocities.

\subsection{Evolution of structure}

We now measure the evolution of spatial structure in the simulations, as quantified by the $\mathcal{Q}$-parameter \citep{Cartwright04,Cartwright09b,Lomax11}, which we define in Eqn.~\ref{qpar_eqn}. A low $\mathcal{Q}$-parameter ($\mathcal{Q} < 0.8$) indicates that substructure is still present, which in turn means that a region is dynamically young, because encounters between stars erase the substructure and push $\mathcal{Q}$ into the smooth regime ($\mathcal{Q}> 0.8$).

\begin{figure*}
  \begin{center}
    \setlength{\subfigcapskip}{10pt}
    \hspace*{-1.5cm} \subfigure[Run J]{\label{N_Qpar-a}\rotatebox{270}{\includegraphics[scale=0.2875]{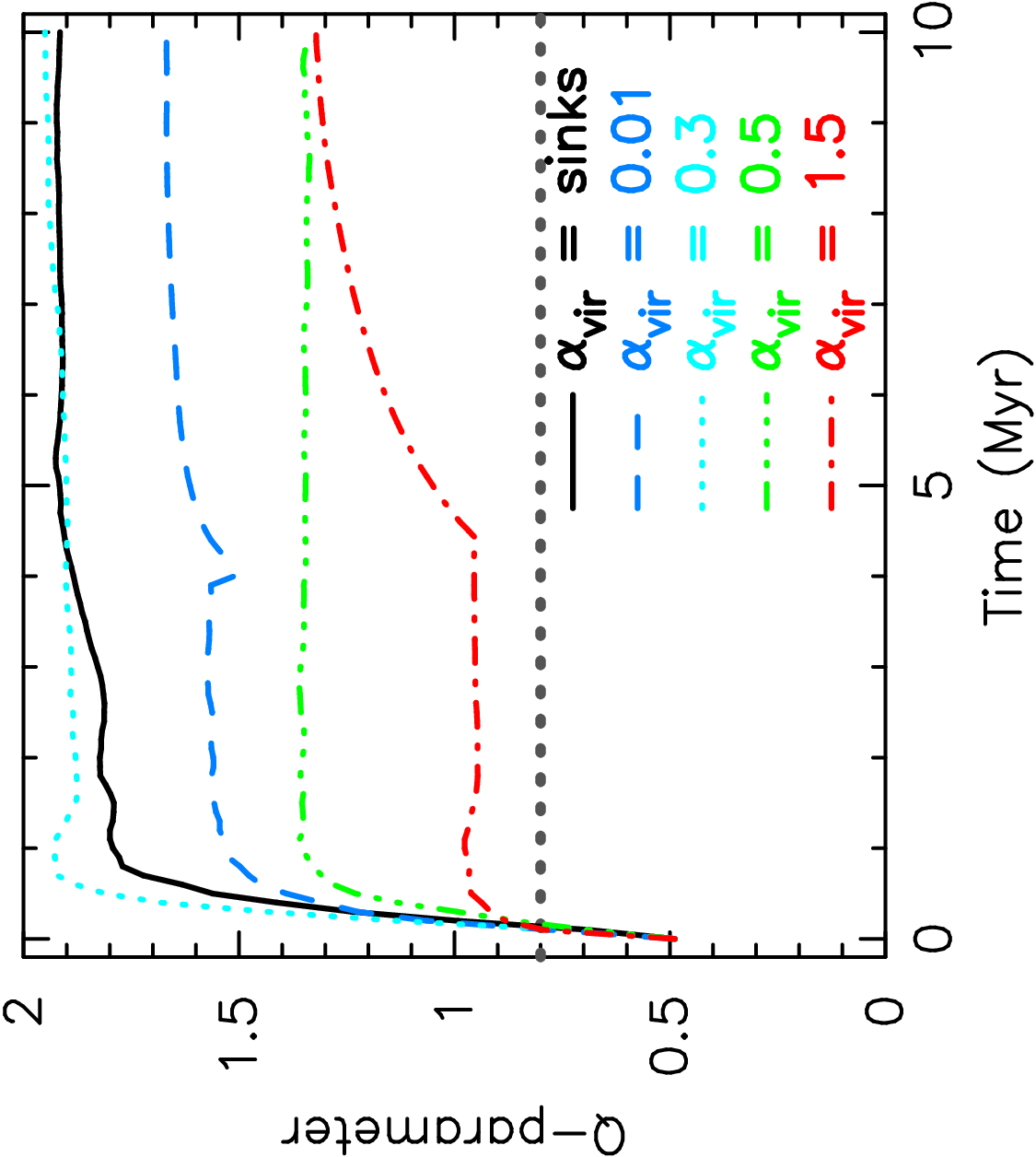}}}
    \hspace*{0.1cm} \subfigure[Run I]{\label{N_Qpar-b}\rotatebox{270}{\includegraphics[scale=0.2875]{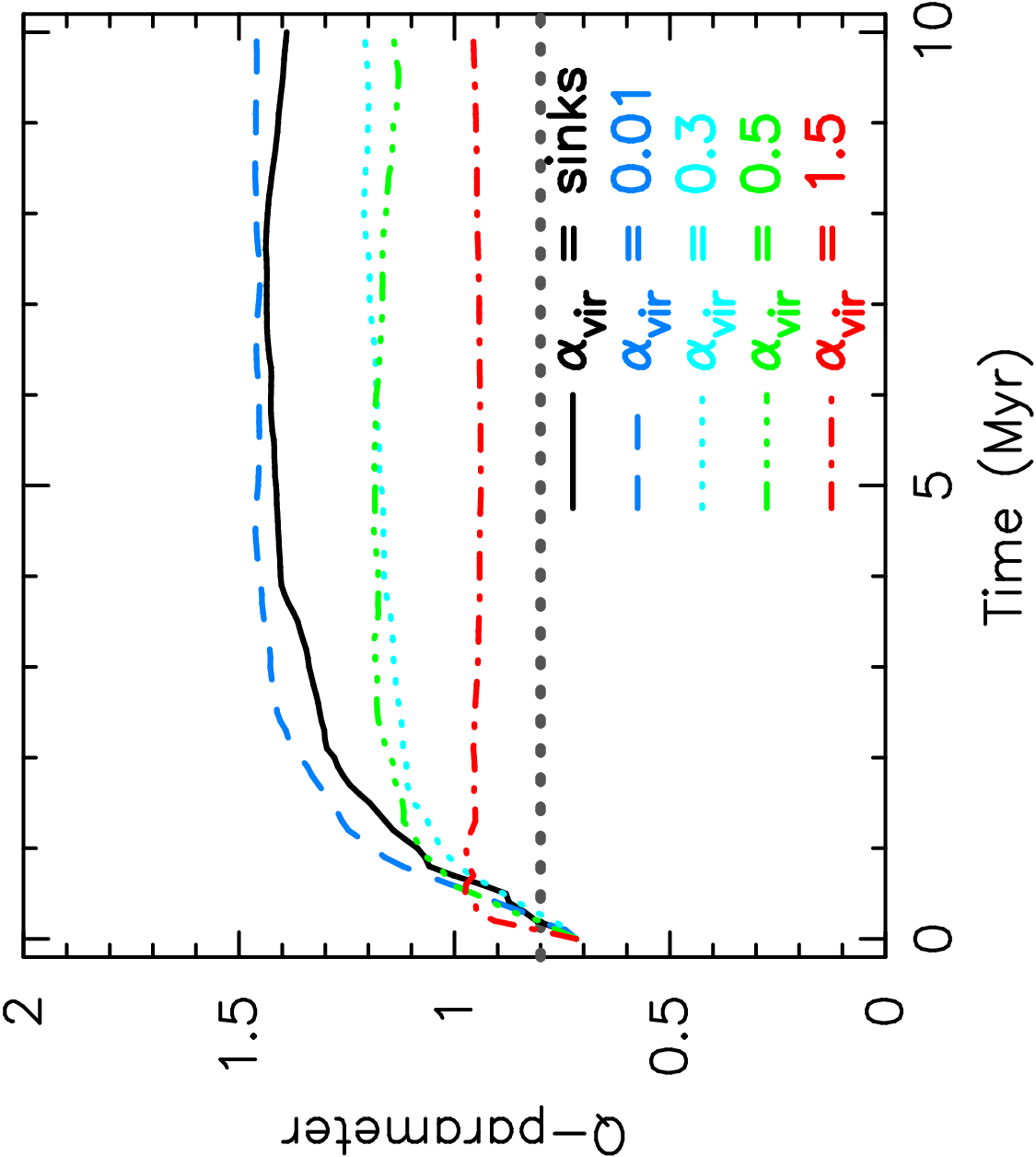}}}
    \hspace*{0.1cm} \subfigure[Run UF]{\label{N_Qpar-c}\rotatebox{270}{\includegraphics[scale=0.2875]{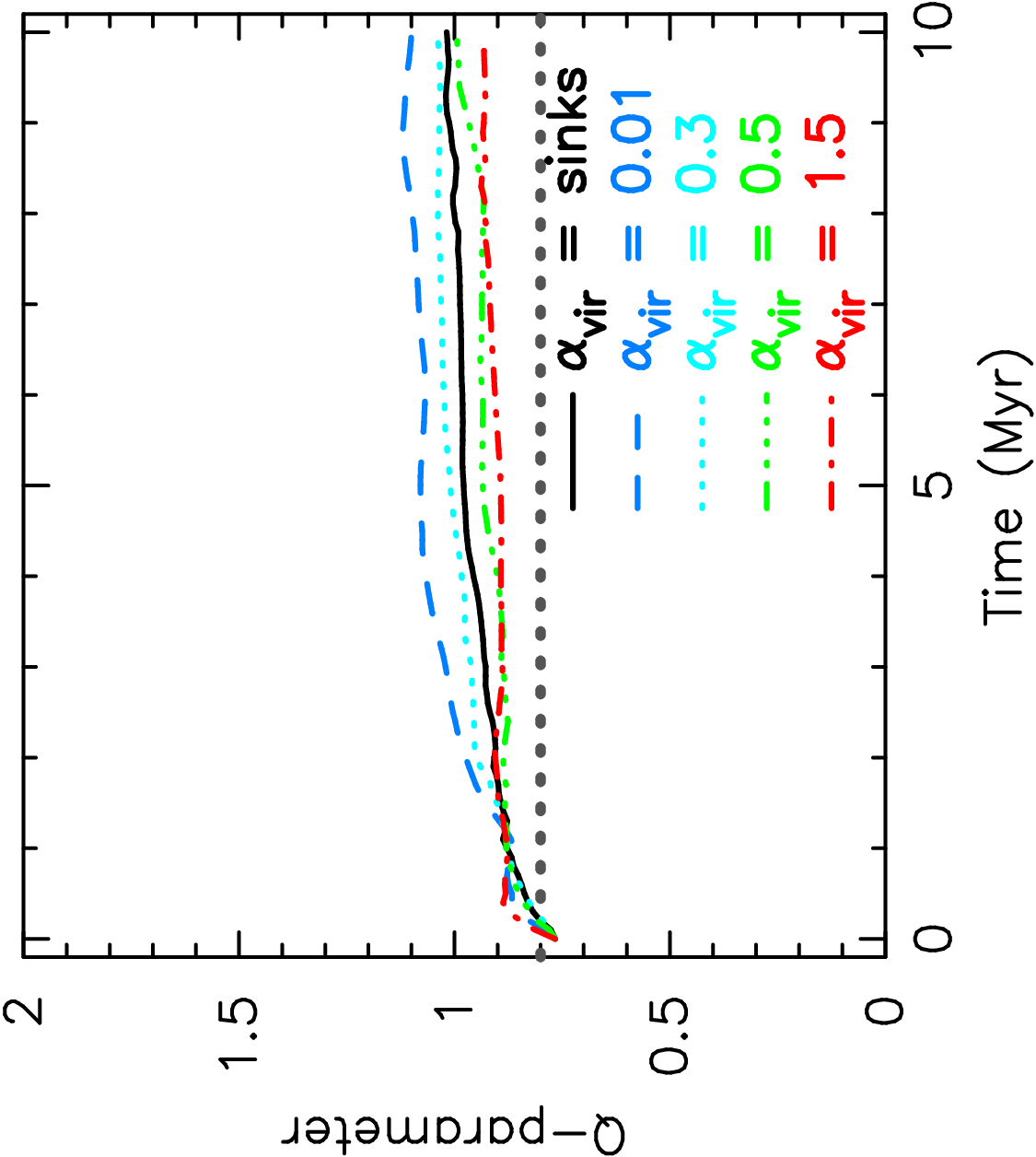}}}
   
\caption[bf]{The evolution of structure (as quantified by the $\mathcal{Q}$-parameter) with time in $N$-body simulations where the stars form in a control run SPH simulation without feedback present. The different lines correspond to different initial virial ratios (i.e.\,\,we run five versions of the same initial conditions, with the only difference being the scaling of the stellar velocities). The horizontal dotted line shows the boundary between a substructured $\mathcal{Q} < 0.8$ and smooth $\mathcal{Q} > 0.8$ regime. }
\label{N_Qpar}
  \end{center}
\end{figure*}

In Fig.~\ref{N_Qpar} we show the evolution of the $\mathcal{Q}$-parameter over time in the $N$-body simulations where the starting initial conditions are the sink particle distributions that formed without feedback.

\begin{figure*}
  \begin{center}
    \setlength{\subfigcapskip}{10pt}
    \hspace*{-1.5cm} \subfigure[Run J]{\label{B_Qpar-a}\rotatebox{270}{\includegraphics[scale=0.2875]{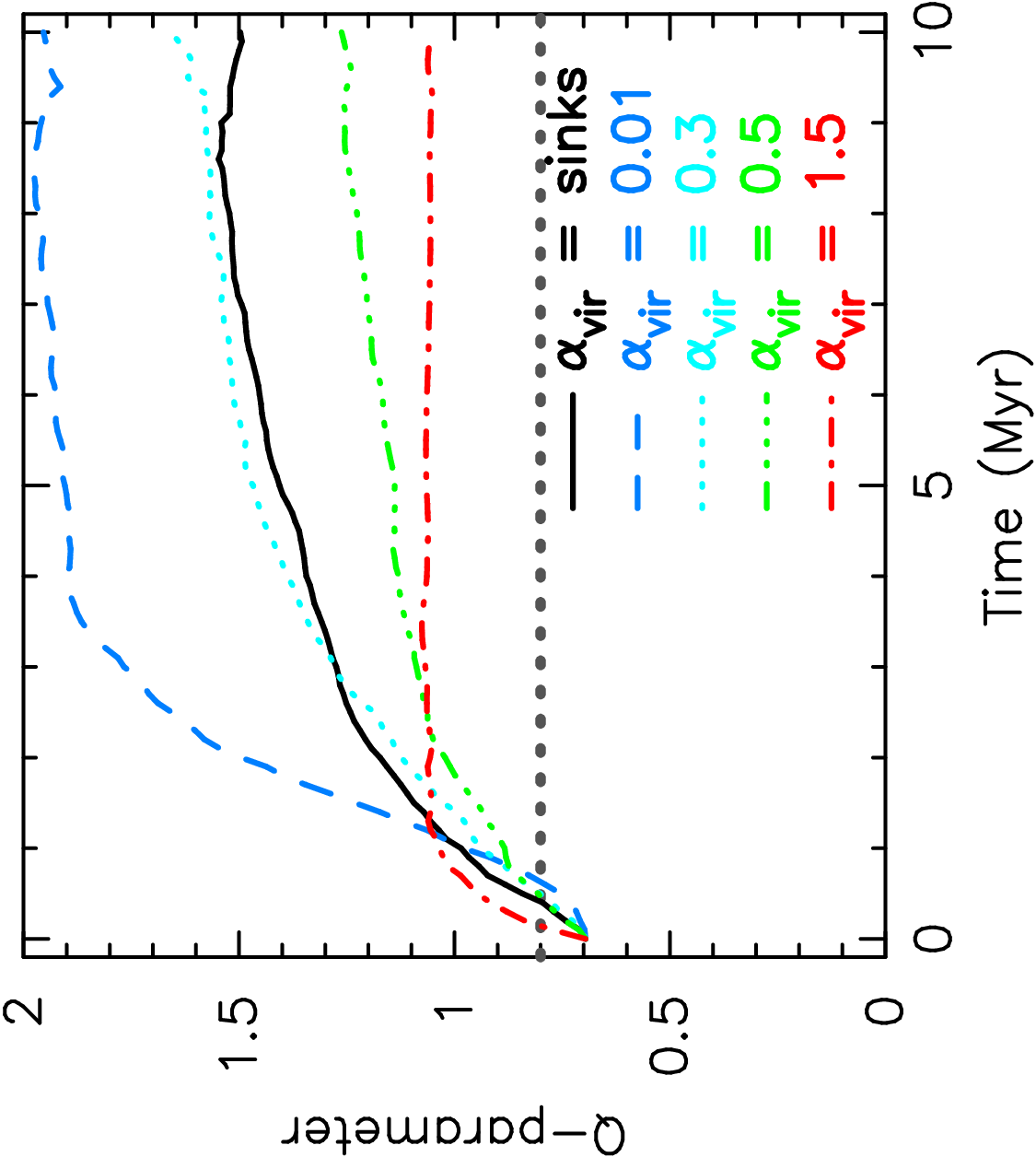}}}
    \hspace*{0.1cm} \subfigure[Run I]{\label{B_Qpar-b}\rotatebox{270}{\includegraphics[scale=0.2875]{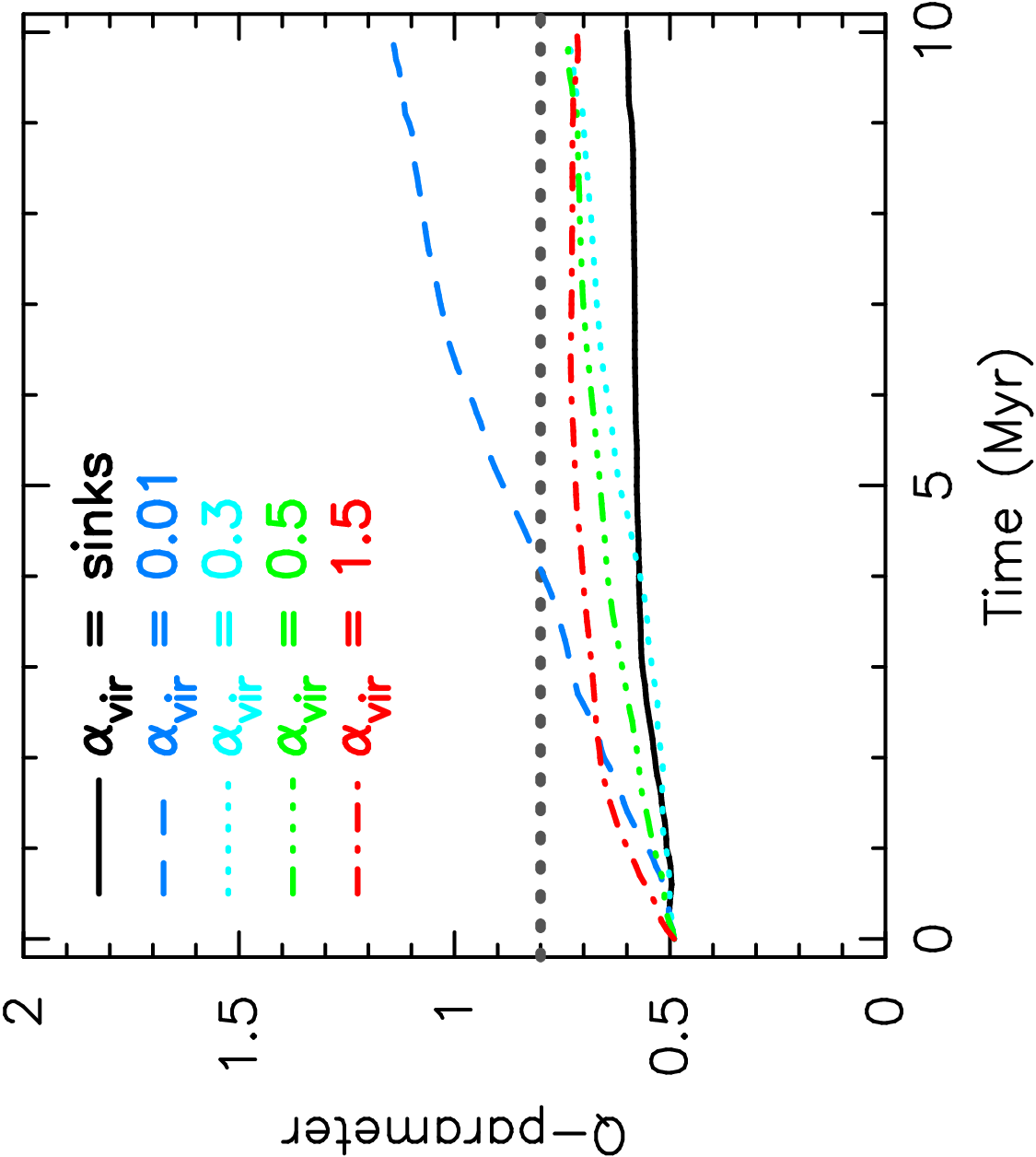}}}
    \hspace*{0.1cm} \subfigure[Run UF]{\label{B_Qpar-c}\rotatebox{270}{\includegraphics[scale=0.2875]{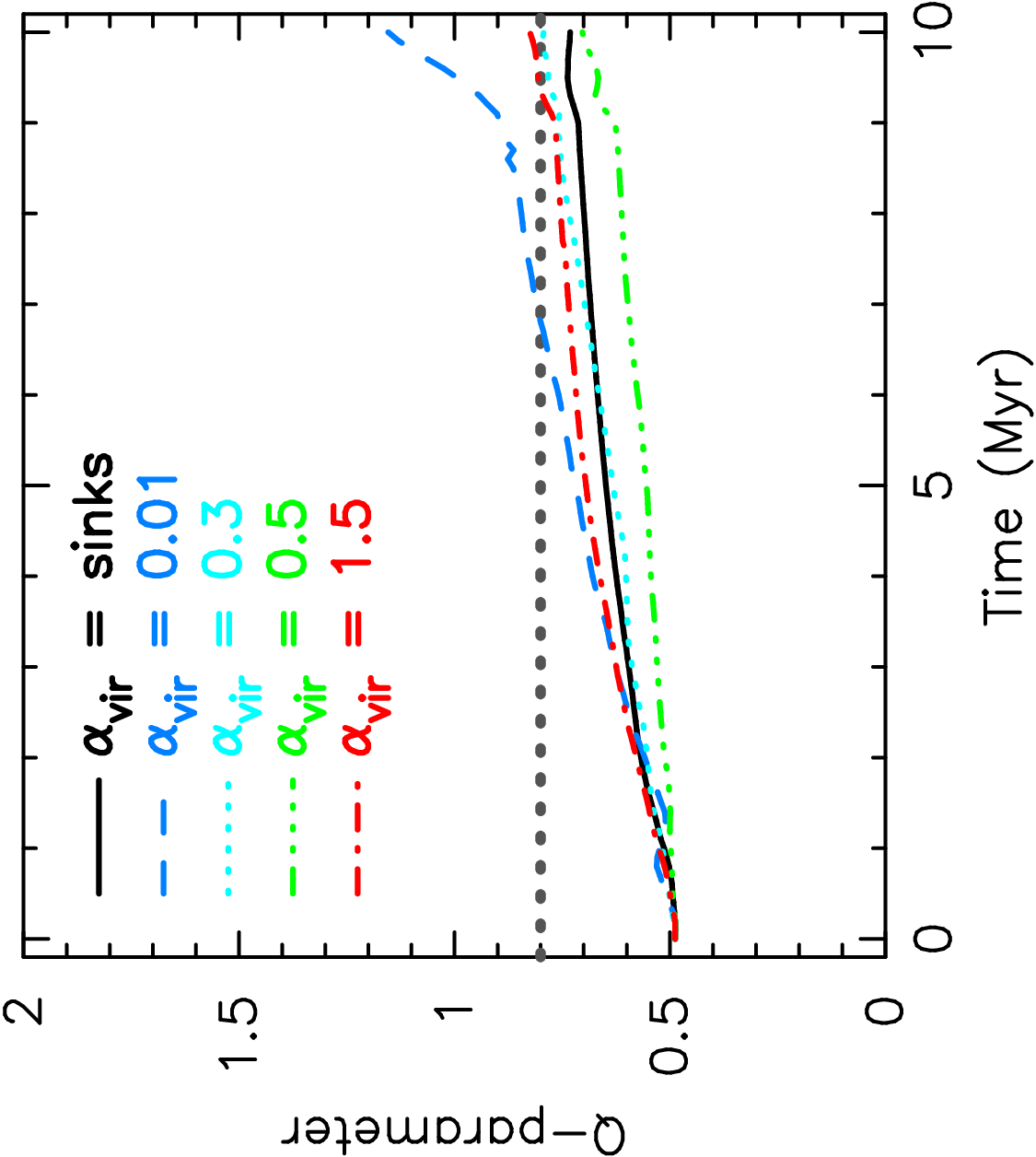}}}

\caption[bf]{The evolution of structure (as quantified by the $\mathcal{Q}$-parameter) with time for $N$-body simulations of stars that formed in the SPH simulations with feedback from photoionising radiation and stellar winds. The different lines correspond to different initial virial ratios (i.e.\,\,we run five versions of the same initial conditions, with the only difference being the scaling of the stellar velocities).  The horizontal dotted line shows the boundary between a substructured $\mathcal{Q} < 0.8$ and smooth $\mathcal{Q} > 0.8$ regime.}
\label{B_Qpar}
  \end{center}
\end{figure*}

In all of these simulations, the initial stellar density is high enough that the substructure is erased on very short ($<$1\,Myr) timescales. This is even the case for the initially supervirial simulations (the red dot-dashed lines), where the stars interact with each other before the overall velocity field (i.e.\,\,virial ratio) causes them to move apart from each other.

As expected from previous studies \citep[e.g.][]{Parker12d,Parker14b,Parker14e}, a more subvirial star-forming region will undergo violent relaxation \citep{LyndenBell67} to a higher degree than a region closer to virial equilibrium \citep{McMillan07,Allison11}, and will therefore form a more centrally concentrated star cluster. This is evident in Figs.~\ref{N_Qpar-a}--\ref{N_Qpar-c}, where the simulations with the lowest initial virial ratio ($\alpha_{\rm vir} = 0.01$, the blue dashed lines) attain the highest $\mathcal{Q}$-parameters.

The simulations in which the stars inherit the velocities of the sink particles from the SPH calculations (the solid black lines) evolve according to which virial ratio the  initial velocity field is closest to. Given that these simulations are predicted to be closest to a scenario where the region undergoes rapid expansion due to sudden mass loss, it is striking that they behave more like a (sub)virial $N$-body simulation.  The reason for this is that -- despite in the case of Run I, a low star-formation efficiency of 12\,per cent which would be expected to facilitate supervirial expansion -- the virial ratio of the sink particles once the gas is removed is subvirial ($\alpha_{\rm vir} = 0.3$).

In the simulations in which the stars formed under the influence of feedback, the stellar densities are lower and therefore the relaxation times (both from the initial violent relaxation if the regions are out of virial equilibrium, and the subsequent two-body relaxation) are longer. When the relaxation times are longer, the structure takes longer to erase, and this is evident in the simulations with feedback which retain substructure throughout the subsequent $N$-body evolution (Figs.~\ref{B_Qpar-b}~and~\ref{B_Qpar-c}). From examination of the $\mathcal{Q}$-parameter in isolation, it is not possible to determine whether the retention of substructure is really due to the regions being supervirial  and the subclusters are moving away form one another so the region remains substructured   (as is the case in the dual feedback Run UF -- Fig.~\ref{B_Qpar-c}) or if it is just due to the low initial stellar densities, meaning it takes several Myr for the substructure to be erased  (as in the case in the dual feedback Run I -- Figs.~\ref{B_Qpar-b}).

\subsection{Evolution of mass segregation}

\begin{figure*}
  \begin{center}
    \setlength{\subfigcapskip}{10pt}
    \hspace*{-1.5cm} \subfigure[Run J]{\label{N_Lambda_10-a}\rotatebox{270}{\includegraphics[scale=0.2875]{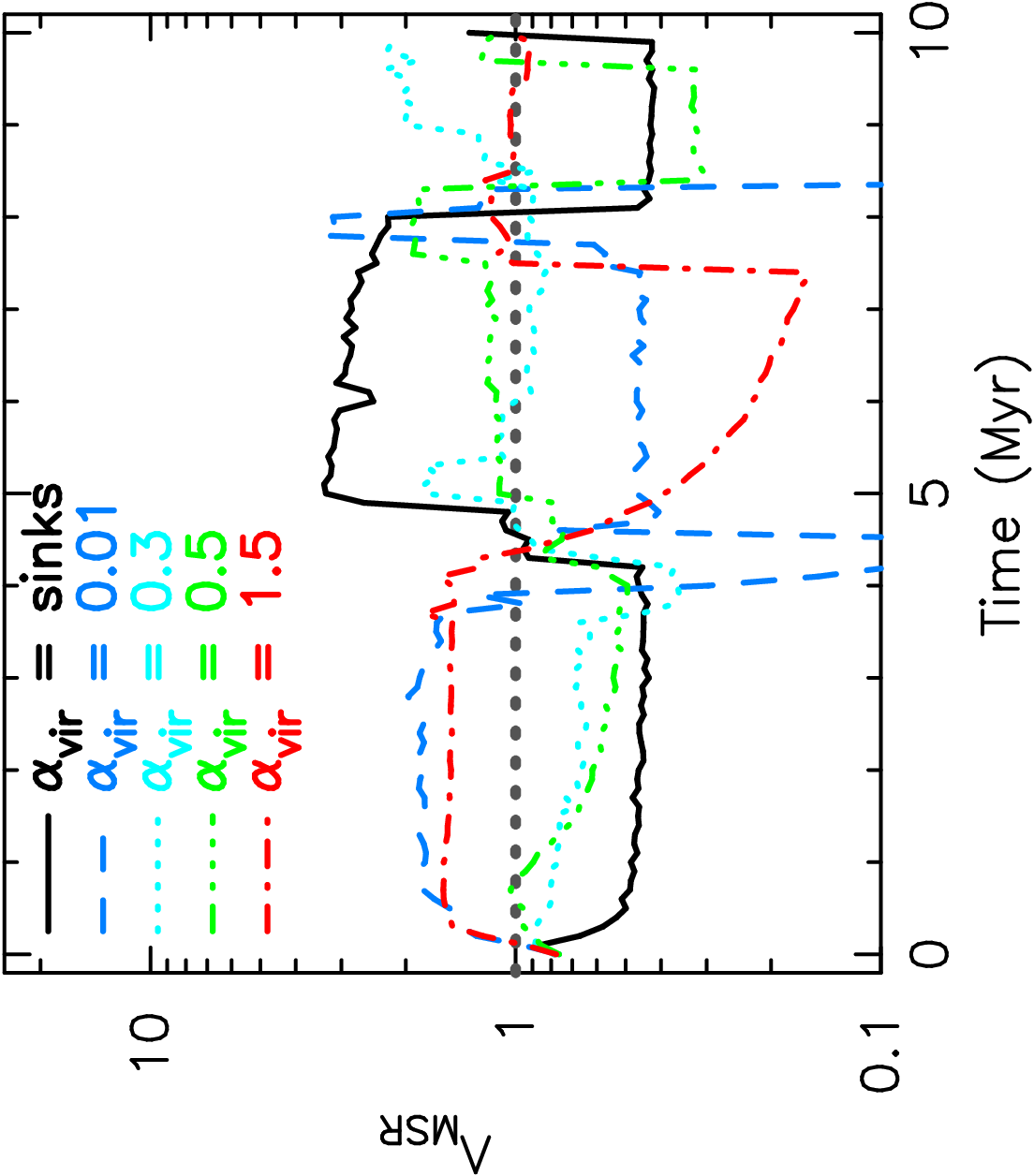}}}
    \hspace*{-0cm} \subfigure[Run I]{\label{N_Lambda_10-b}\rotatebox{270}{\includegraphics[scale=0.2875]{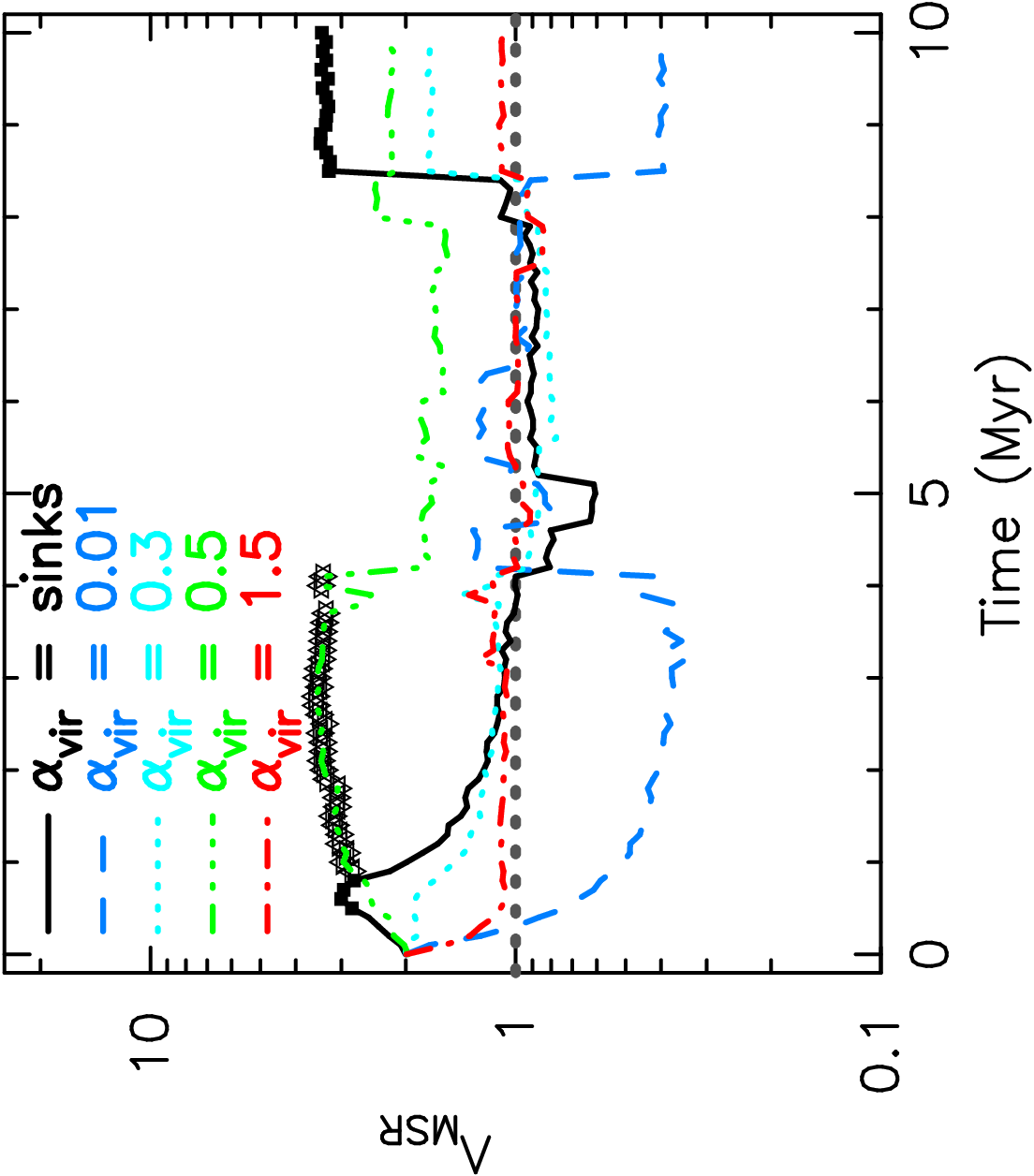}}}
    \hspace*{-0cm} \subfigure[Run UF]{\label{N_Lambda_10-c}\rotatebox{270}{\includegraphics[scale=0.2875]{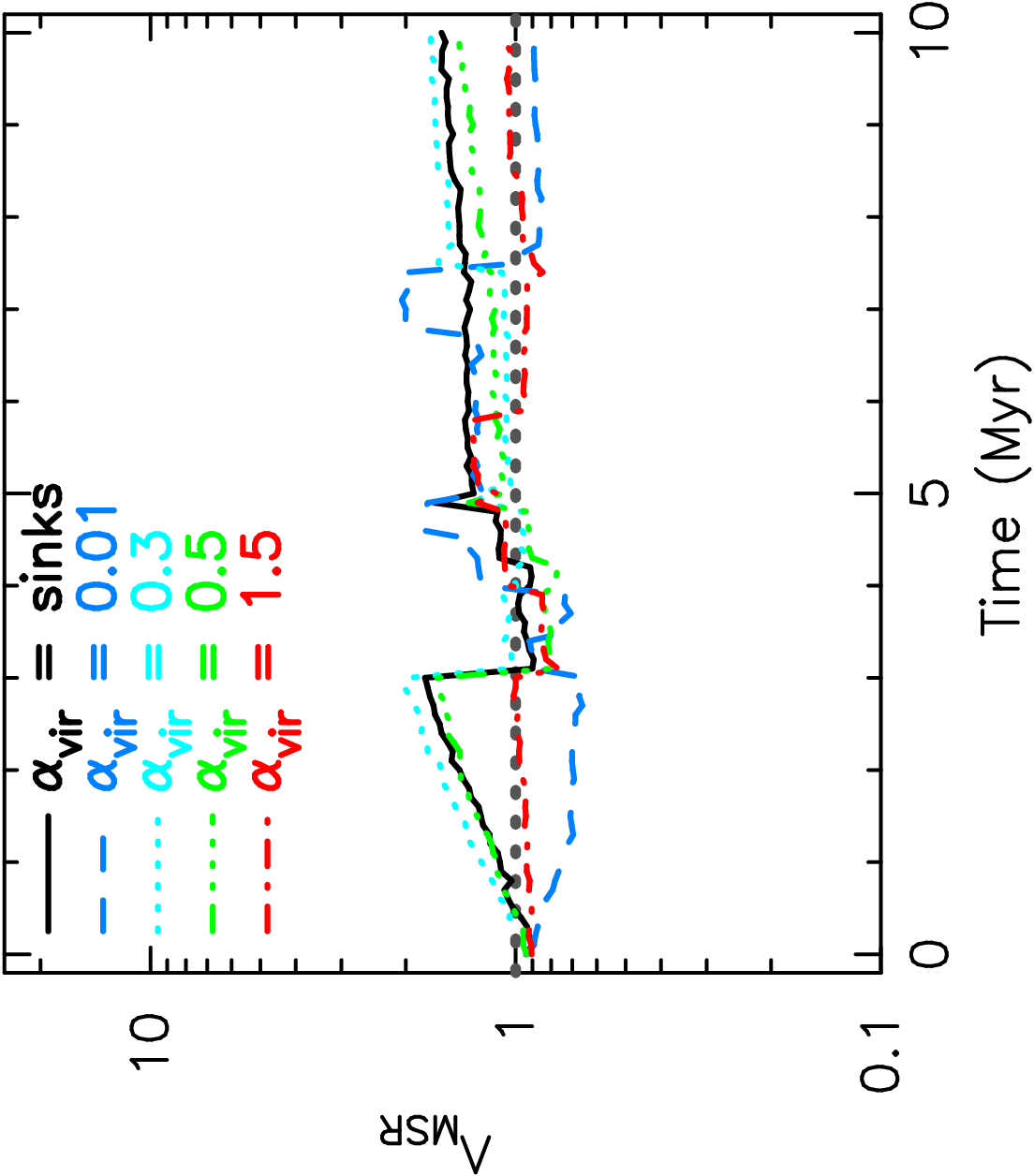}}}


\caption[bf]{The evolution of the $\Lambda_{\rm MSR}$ mass segregation ratio with time in $N$-body simulations where the stars form in a control run SPH simulation without feedback present. The different lines correspond to different initial virial ratios (i.e.\,\,we run five versions of the same initial conditions, with the only difference being the scaling of the stellar velocities). Where $\Lambda_{\rm MSR} >> 1$, we plot a filled symbol. The horizontal dotted line shows $\Lambda_{\rm MSR} = 1$, i.e.\,\,no mass segregation.}
\label{N_Lambda_10}
  \end{center}
\end{figure*}

\begin{figure*}
  \begin{center}
    \setlength{\subfigcapskip}{10pt}
    \hspace*{-1.5cm} \subfigure[Run J]{\label{B_Lambda_10-a}\rotatebox{270}{\includegraphics[scale=0.2875]{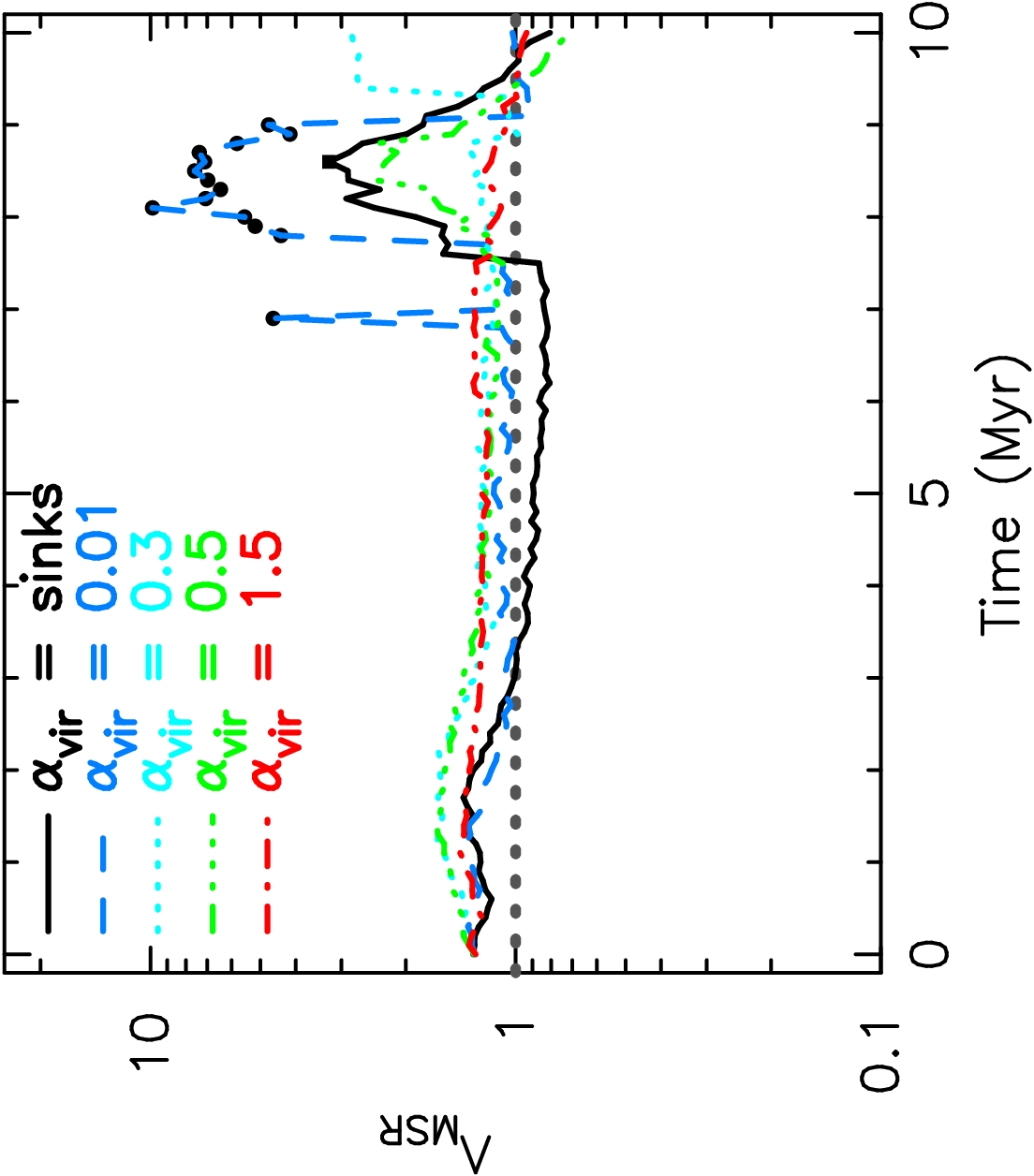}}}
    \hspace*{0.1cm} \subfigure[Run I]{\label{B_Lambda_10-b}\rotatebox{270}{\includegraphics[scale=0.2875]{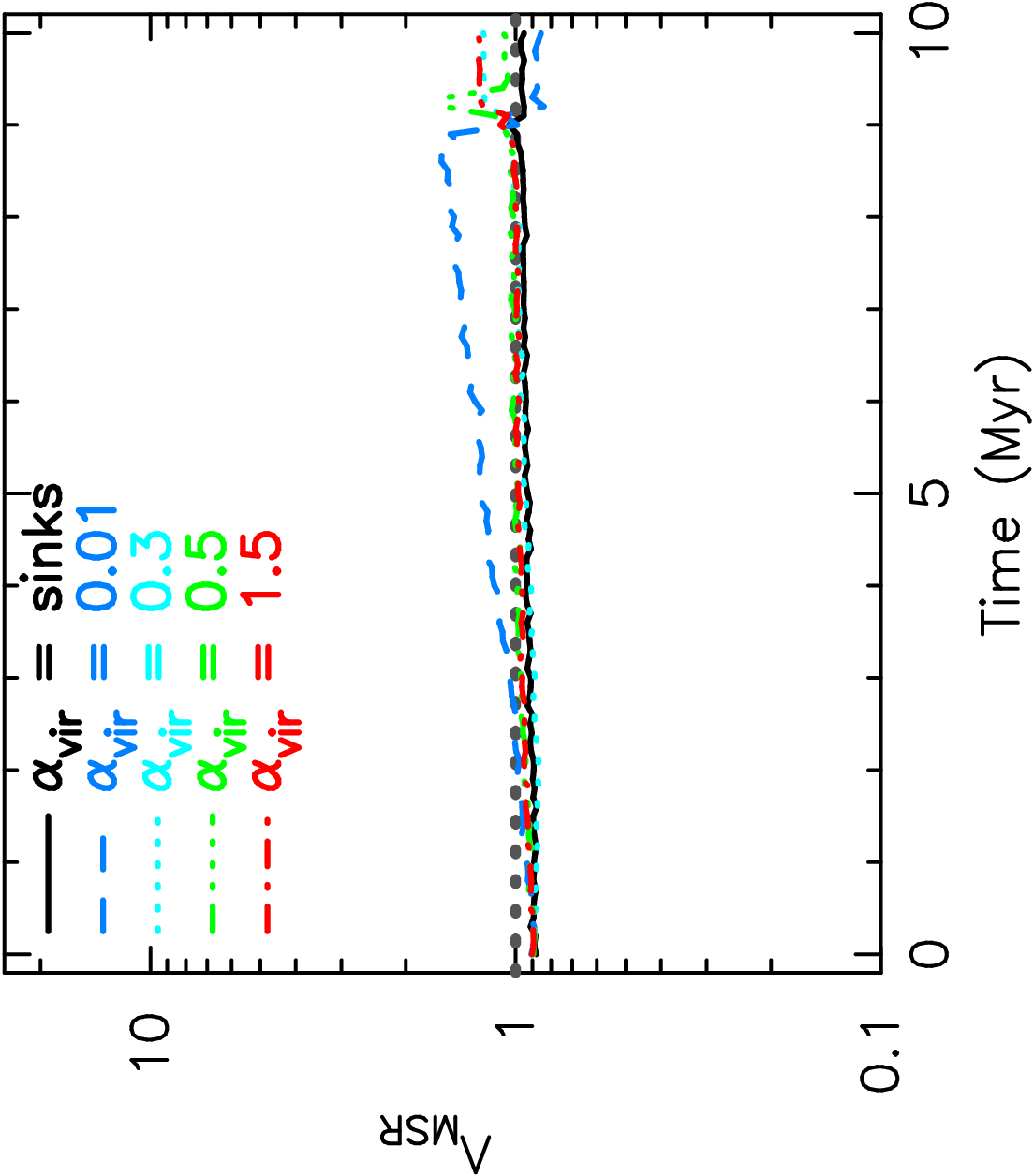}}}
    \hspace*{0.1cm} \subfigure[Run UF]{\label{B_Lambda_10-c}\rotatebox{270}{\includegraphics[scale=0.2875]{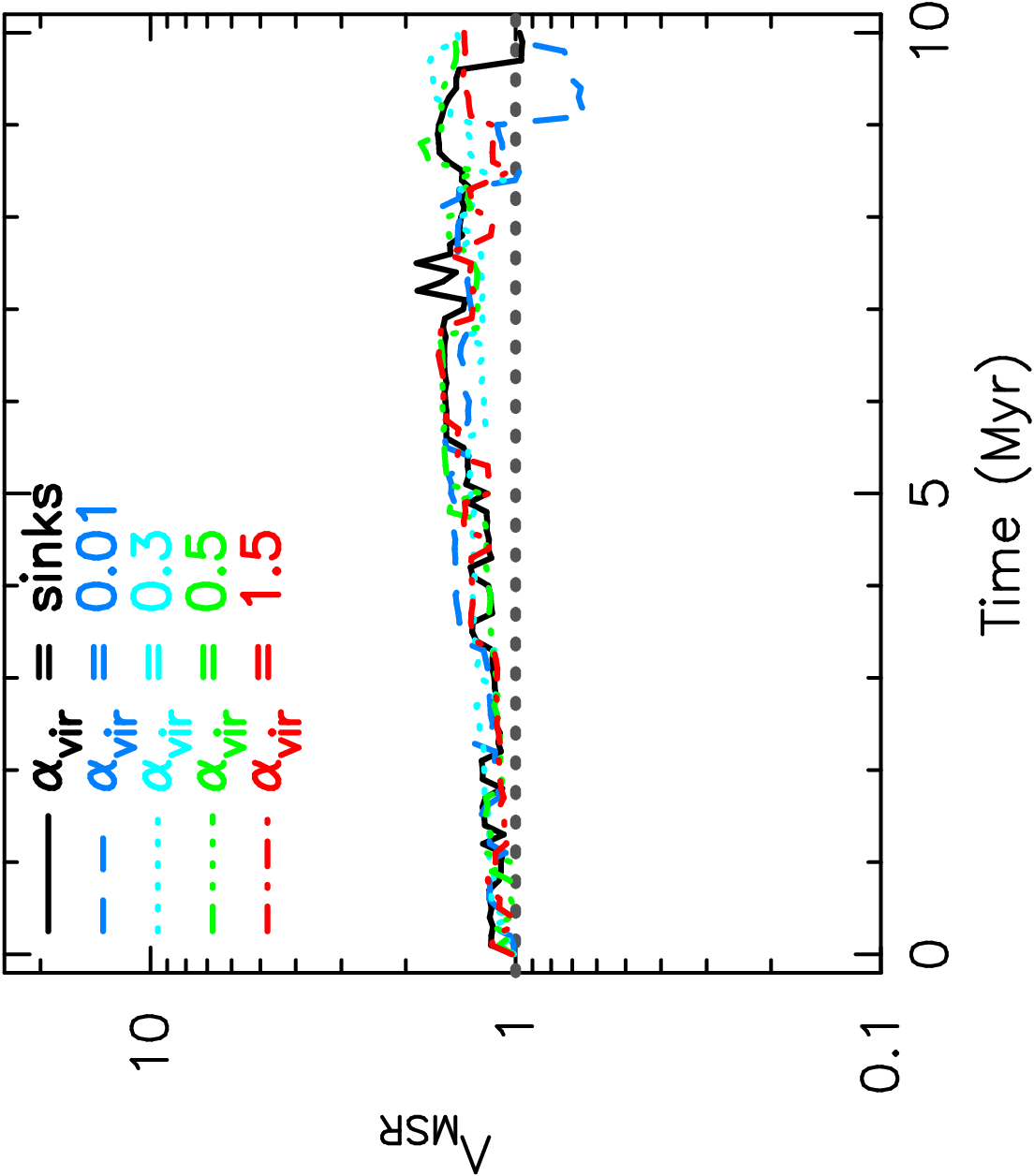}}}
\caption[bf]{The evolution of the $\Lambda_{\rm MSR}$ mass segregation ratio with time for $N$-body simulations of stars that formed in the SPH simulations with feedback from photoionising radiation and stellar winds. The different lines correspond to different initial virial ratios (i.e.\,\,we run five versions of the same initial conditions, with the only difference being the scaling of the stellar velocities). Where $\Lambda_{\rm MSR} >> 1$, we plot a filled symbol.  The horizontal dotted line shows $\Lambda_{\rm MSR} = 1$, i.e.\,\,no mass segregation. }
\label{B_Lambda_10}
  \end{center}
\end{figure*}

Mass segregation -- usually defined as the over-concentration of the most massive stars in the central locations of a star-forming region -- has long been thought to be the outcome of `competitive accretion' \citep{Zinnecker82,Bonnell97,Bonnell98,Bonnell98b,Bonnell01,Bonnell06,Maschberger11}, whereby the stars with access to the largest gas reservoirs -- assumed to be in central locations in a GMC -- grow to the largest masses. Whilst this behaviour is seen in \emph{some} purely hydrodynamical simulations of star formation, the addition of feedback processes severely limits the gas available, and therefore `primordial' mass segregation does not usually arise from the star formation process in simulations  \citep{Parker15a,Parker17b,Parker24}. There is some observational evidence that pre-stellar cores are mass segregated in some regions \citep{Plunkett18,Konyves20} but not in others \citep{Parker18a}.  \citet{Alcock19} show that any observed mass segregation in cores would translate into mass segregation in the stars, unless the stars moved far from their birth cores' locations or underwent a very contrived fragmentation process. 

The lack of mass segregation in the majority of our simulations is similar to the more recent STARFORGE simulations \citep{Guszejnov22} who find values for $\Lambda_{\rm MSR} <2$, which \citet{Parker15b} show can occur through random fluctuations in distributions of only a few hundred objects, and is not an indication of true mass segregation. \citet{Guszejnov22} also define mass segregation using using an offset-ratio method, which quantifies how far massive stars are away from a (sub)cluster centre. Unfortunately, \citet{Parker15b} demonstrated that the offset-ratio method is flawed and does not reliably quantify mass segregation.  That mass segregation is observed in cores only in some regions, and occurs only occasionally in simulations, suggests it is not an expected outcome of the star formation process.

Instead, mass segregation can occur through dynamical relaxation, and usually the stellar density must be high ($>10^3$\,M$_\odot$\,pc$^{-3}$) to facilitate dynamical mass segregation within several Myrs \citep{McMillan07,Allison09b,Allison10}. A notable feature of dynamical mass segreation is that it can be very transient. Massive stars form their own dynamical sub-system within a star-forming region \citep{Allison11} and this can lead to ejections of one or more of the massive stars, which would decrease the degree of mass segregation \citep{Parker16c} (though possibly only temporarily, until the errant massive star(s) are replaced in the subsystem).

We see a small degree of dynamical mass segregation in the $N$-body evolution of our star-forming regions, most notably in the simulations in which the stars form without feedback (Fig.~\ref{N_Lambda_10}). We note that the mass segregation here is not because these simulations form more massive stars (and have a top-heavy IMF) -- the $\Lambda_{\rm MSR}$ ratio is a relative measure -- but rather because they form with higher stellar densities, which facilitates a faster dynamical relaxation.

In the simulations in which the stars formed without feedback, significant mass segregation occurs in two of the realisations of Run I (Fig.~\ref{N_Lambda_10-b}), but in neither does it remain significant for the duration of the simulation.  Indeed, in the simulation where the stars inherit the sink particle velocities (the solid black line),  mass segregation  occurs around 1\,Myr, disappears almost immediately, before re-occurring after 8.5\,Myr, most likely due to the stellar evolution (mass loss) of the most massive stars, which then changes the membership of the most massive subset.

In Run J (Fig.~\ref{N_Lambda_10-a}) and Run UF (Fig.~\ref{N_Lambda_10-c}) there is no significant mass segregation.

In the simulations where the stars form under the influence of feedback, the later dynamical evolution rarely leads to dynamical mass segregation because of the relatively low stellar densities, meaning that dynamical mass segregation does not occur. An exception to this is in the evolution of Run J when the velocities of the stars are set to be very subvirial ($\alpha_{\rm vir} = 0.01$, the blue dashed line in Fig.~\ref{B_Lambda_10-a}). This late-stage mass segregation (which occurs between 7 -- 9 \,Myr) does not happen in either Run I (Fig.~\ref{B_Lambda_10-b}) or Run UF (Fig.~\ref{B_Lambda_10-c}).

The simulations we present here are the first in which mass segregation has been quantified in different runs of the same simulations, where the positions of the stars and their masses are constant, but the velocities are different. We find very little difference between the different initial velocity fields, even between the very extremes ($\alpha_{\rm vir} = 0.01$ versus $\alpha_{\rm vir} = 1.5$).

\section{Discussion}
\label{discuss}

Our primary result is that instantaneously removing the leftover gas from simulations of star formation where significant feedback is present (stellar winds and photoionisation) does not produce a supervirial velocity distribution in the stars. The long-term evolution of the stars in an $N$-body simulation is much more similar to a virialised, or even subvirial star-forming region, rather than a region that is very supervirial.

One interpretation of this could be that the stars' velocities are reacting to the feedback (and the removal of gas) before we remove the remaining gas potential that has not formed stars. The stars are therefore adjusting to a gradual change in the background potential before the remaining potential is removed. However, we see the same behaviour of the velocities in the simulations where stars form without feedback, and where there is still a significant gas potential in the locations where the stars have formed. In these simulations, the removal of the gas would in theory cause a rapid dispersal of the stars. This in face does happen in one set of simulations without feedback -- Run UF -- but removing the gas in the remainder of the simulations does not have the same effect.

An interpretation that would explain the evolution of both the simulations with, and without, feedback, is that the star formation efficiency is high enough to prevent the dipsersal of the region, if the stars themselves have subvirial velocities. In the classical picture of gas expulsion, a star formation efficiency of less than $\sim$30\,per cent would cause a star-forming region to disrupt after sudden gas removal \citep{Goodwin06,Parker17d}. However, the important parameter is the \emph{effective} star formation efficiency \citep{Goodwin09}, which takes into account the virial ratio of the stars. If the stars have subvirial velocities, then the true star formation efficiency (the amount of gas converted into stars) can be much lower than 30\,per cent and the region can still remain bound.

Dividing the total mass of stars, $M_{\rm region}$, by the total cloud mass $M_{\rm cloud}$, in Table~\ref{cluster_props}, suggests true star formation efficiencies of between 3 and 32\,per cent, but the majority of these regions have virial ratios significantly lower than $\alpha_{\rm vir}^{\rm stars} =  0.5$ (virial equilibrium).  Therefore, the effective star formation efficiency will be much higher \citep{Goodwin09}, and this prevents even the simulations with significant gas remaining from disrupting, once the gas is (instantaneously) removed.

The majority of our simulations where the $N$-body simulations inherit the sink particle velocities do not evolve in as drastic a manner as the highly supervirial simulations (the red dot-dashed lines in the figures), which suggests that if a star-forming region is affected by gas removal, it is unlikely to be as unbound as these highly supervirial simulations.

We note, however, that even a (sub)virial star-forming region will expand due to a combination of violent, and then two-body relaxation, and the degree to which the region will expand is usually governed by how densely packed the stars are prior to the $N$-body evolution.

Furthermore, the expansion of a star-forming region following violent relaxation can often appear to be supervirial when looking at the velocity dispersion of stars \citep[even when corrected for binary motion,][]{Cottaar12b}, as the velocity dispersion is ``frozen in'' from its peak value during the violent relaxation \citep{Parker16b}.

A significant amount of observational effort has been invested in measuring the structure and degree of mass segregation in star-forming regions. However, the majority of observed star-forming regions have $\mathcal{Q}$ parameters in the range 0.7 - 0.9 \citep{Hetem15,Dib18,DaffernPowell20}, which indicates an abiguous structure (neither substructured, nor significantly centrally concentrated) and most regions do not display significant mass segregation \citep{Parker17a,Dib18}. Other measures of the spatial distribution \citep[e.g.][]{Alfaro16,Gonzalez17,Jaffa17,Arnold19,Buckner19,BlaylockSquibbs22} are similarly inconclusive. \citet{Hetem15} provide a direct comparison of observed cluster structure to similar simulations to those presented here \citep{Parker13a}, and find no observational smoking gun for a particular set of initial conditions, either with, or without feedback.

We highlight several caveats with our work. First, the SPH simulations may not be representative of star formation, either because they happen to be statistical outliers, or that they include or exclude certain physics. For example, more recent simulations of star formation with feedback have included the effects of magnetic fields during the formation, and added jets to the feedback mechanisms, in addition to winds, supernovae and photoionising radiation \citep{Grudic21}. Our simulations have a somewhat narrow initial conditions range, all being Solar metallicity with similar cloud masses, although we do vary the initial cloud radii and the virial ratios of the clouds.

Second, the SPH simulations did not model the effects of supernovae of the most massive stars, which in principle could remove even more of the background gas potential ealier on. However, we note that the latest stellar evolution models \citep{Ertl16,Limongi18} suggest that stars $>$25\,M$_\odot$ are more likely to collapse to a black hole without exploding as a supernova, and that stars with lower masses do not explode until after 10\,Myr, due to rotation effects. Furthermore, the presence of binaries, and their evolution (which is included in the $N$-body calculations, but not the parent SPH calculations) can affect feedback timescales \citep{deMink13,Gotberg19}, especially if the majority of massive stars form in binary or multiple systems \citep{Sana13,Sana25}. 

Thirdly, because we take the positions, masses and velocities from the SPH simulations, we cannot vary these distributions on a statistical level, i.e.\,\,we cannot run different versions of the same $N$-body simulation. Therefore, the $N$-body evolution could in principle be an outlier, though we note that the same behaviour occurs in  almost all of the $N$-body evolution of the stars taken from the different SPH simulations.

\section{Conclusions}
\label{conclude}

We take the distribution of stars (sink particles) from SPH simulations of star formation and then evolve them for a further 10\,Myr as $N$-body simulations to determine how much of an effect the removal of gas left over from star formation has a the long-term evolution of a star-forming region. Our conclusions are as follows.\\

(i) Simulations in which the stars inherit the velocities of the sink particles in the SPH simulation evolve in a similar manner to other realisations where the velocities are scaled to be (sub)virial. Only one simulation with the directly inherited velocities evolves in a similar way to a highly supervirial simulation.

(ii) We might expect the sudden removal of gas in the control-run  simulations where feedback does not act on the GMC to have more of an effect on the later dynamical evolution of the stars. However, this is not the case, and any differences in the dynamical evolution of the control runs  compared to the feedback runs are due to the higher stellar densities of the control runs after all of the stars have formed.

(iii) The largest difference in the evolution of the simulations with different velocity fields is seen in the $\mathcal{Q}$-parameter, which quantifies the amount of spatial structure in a distribution of stars. For the more subvirial simulations, the $\mathcal{Q}$-parameter reaches higher values, because the degree of dynamical relaxation is higher. In contrast, the $\mathcal{Q}$ values tend to be much lower for the supervirial simulations (though not always low enough to indicate substructure). On the other hand, the exact velocity field is not distinguishable in the evolution of mass segregation, or the median local surface density of the stars.\\

Taken together, our results suggest that a signature of gas removal from feedback from massive stars is unlikely to be observed in real star-forming regions, and that the dissolution of star-forming regions is more likely to be governed by dynamical relaxation (and the influence of the Galactic tidal field), rather than gas expulsion.

\section*{Acknowledgments}

We thank the anonymous referee for a very helpful review. RJP acknowledges support from the Royal Society in the form of a Dorothy Hodgkin Fellowship.

\section*{Data Availability}

The data used to produce the plots in this article will be shared on reasonable request to the 
corresponding author.

\bibliographystyle{mnras}
\bibliography{general_ref}

\label{lastpage}

\end{document}